\documentclass{emulateapj}
\usepackage{natbib}
\usepackage{amsmath,amsfonts,amssymb,pslatex}
\usepackage{graphicx}

\def\idm#1{{\mbox{\scriptsize #1}}}

\newcommand\Ym{\langle Y\rangle}
\newcommand\Chi{{(\chi^2_\nu)^{1/2}}}
\def\astrobj#1{#1}

\def\url#1{\texttt{#1}}

\newcommand\stara{{\astrobj{HD~202206}}}
\newcommand\starb{{\astrobj{14~Her}}}
\newcommand\starc{{\astrobj{HD~37124}}}
\newcommand\stard{{\astrobj{HD~108874}}}

\begin{document}

%
\title{Orbital configurations and dynamical stability of multi-planet systems
around Sun-like stars HD~202206, 14~Her, HD~37124 and HD~108874}
%

\author{Krzysztof Go\'zdziewski\altaffilmark{1},
        Maciej Konacki\altaffilmark{2}, and
        Andrzej J. Maciejewski\altaffilmark{3}
       }
\altaffiltext{1}{Toru\'n Centre for Astronomy,
  N.~Copernicus University,
  Gagarina 11, 87-100 Toru\'n, Poland; k.gozdziewski@astri.uni.torun.pl}
\altaffiltext{3}{
Nicolaus Copernicus Astronomical Center, Polish Academy of Sciences,
Rabia\'nska 8, 87-100 Toru\'n, Poland, and
Department of Geological and Planetary Sciences, California
Institute of Technology, MS 150-21, Pasadena, CA 91125, USA; 
maciej@ncac.torun.pl}  
\altaffiltext{3}{Institute of Astronomy, University of Zielona G\'ora,
  Podg\'orna 50, 65-246 Zielona G\'ora, Poland; maciejka@astro.ia.uz.zgora.pl}
\begin{abstract}
We perform a dynamical analysis of the recently published radial velocity (RV)
measurements of a few solar type stars which host multiple Jupiter-like planets.
In particular, we re-analyze the data for HD~202206, 14~Her, HD~37124 and
HD~108874.  We derive dynamically stable configurations which  reproduce the 
observed RV signals using  our method called GAMP (an acronym of the Genetic 
Algorithm with MEGNO Penalty). The GAMP relies  on the $N$-body dynamics and
makes use of genetic algorithms merged with  a stability criterion. For this
purpose, we use the maximal Lyapunov exponent  computed with the dynamical fast
indicator MEGNO. Through a dynamical analysis of the phase-space in a 
neighborhood of the obtained best-fit solutions, we derive meaningful limits  on
the parameters of the planets. Without taking into account the stability
criterion and due to narrow observational windows, the orbital elements  of the
outermost planets are barely constrained and both Keplerian,  as well as
Newtonian, best-fit solutions often correspond to self-disrupting 
configurations. We demonstrate that GAMP is especially well suited for the
analysis  of the RV data which only partially cover the longest orbital period
and/or correspond  to multi-planet configurations involved in low-order mean
motion resonances (MMRs).  In particular, our analysis reveals a presence of a
second Jupiter-like planet in the 14~Her system (14~Her~c) involved in a 3:1 or
6:1~MMR with the known  companion~b. We also show that the dynamics of the
HD~202206 system may be  qualitatively different when coplanar and
mutually-inclined orbits of the  companions are considered. We demonstrate that
the two outer planets in the HD~37124 system may reside in a close neighborhood
of the 5:2~MMR. Finally, we found a clear indication that the \stard{} system
may be very  close to, or locked in  an exact 4:1~MMR.
\end{abstract}


\section{Introduction}

Finding the best-fit solutions to radial velocity (RV) observations of stars 
with more than one planet that cover only partially the longest orbital period 
is difficult.  Modeling the RV data with a kinematic superposition of Keplerian 
orbits or even with a full $N$-body dynamics often leads to configurations with 
not well constrained eccentricity of the outermost planetary companion
\citep{Jones2002b,Gozdziewski2003e}. In the statistically optimal best-fit
solutions, the eccentricities can be large and quickly (on the time-scale of 
thousand of years) lead to catastrophic collisional instability. Moreover,  the
validity of a superposition of kinematic Keplerian signals can be very  
problematic for systems involved in low-order mean motion resonances (MMRs). 
Due to significant uncertainties of the best-fit parameters, even the $N$-body 
model of the RV curve that incorporates the mutual gravitational interactions 
frequently yields unstable configurations because the model is "blind'' to the 
sophisticated, fractal-like structure of the orbital parameter space as
predicted by the fundamental Kolmogorov-Arnold theorem  \citep{Arnold1978}. 
According to this theorem, the phase-space of a planetary system is  {\em
discontinuous} with respect to the requirement of stability.

An ideal fitting algorithm should find a solution which reproduces the RV data 
and simultaneously corresponds to a stable planetary configuration. The most 
frequently used notion of the term {\em stable} means {\em not disrupting or 
qualitatively changing during short periods of time, say million of years}. 
This idea has been already used by many authors modeling the RV data. Our  first
attempt to use this idea resulted in a dynamical confirmation of the 2:1 MMR of
in the HD~82943 system \citep{Gozdziewski2001b}. Recently, 
\cite{FerrazMello2005}, \cite{Correia2005} and \cite{Vogt2005} have applied 
such an approach in the analysis of the RV data of multi-planet systems (in 
particular, of hypothetically resonant configuraitons). Often, a stability
criterion  is applied {\em after} the mathematically best-fit solution is found
and the orbital elements are  then adjusted to obtain a stable configuration. We
further show that such  a modification of the  best-fit initial condition does
not necessarily give an optimal  solution.

In our relatively new method called the Genetic Algorithm with MEGNO Penalty 
(GAMP), the stability analysis is an {\em internal} part of the fitting 
procedure \citep{Gozdziewski2003e,Gozdziewski2005a}. We treat the dynamical
behavior in terms of the chaotic and regular (or weakly-chaotic) states as an
additional observable at the same level of importance as the RV measurements
are. The unstable solutions are penalized by artificially increasing their
$\Chi$. For determining the character of motions, we rely on the computation of
the maximal Lyapunov exponent through the MEGNO indicator
\citep{Cincotta2000,Cincotta2003,Cincotta2004}. Apparently, the use of such a
formal criterion  of the stability for modeling real data may be problematic.
Almost any planetary system, including our own, can be very close to a chaotic
state. Nevertheless, we expect that even if chaos appears, it should not impair
the astronomical stability  \citep{Lissauer1999} meaning that planetary  orbits
are  bounded over a very long time and any collisions or ejection of planets do
not occur. However, for configurations involving Jupiter-like companions in
close orbits with large eccentricities, the formal stability seems to be well
related to the astronomical stability. A serious  complication is that there is
not known any general relation between the Lyapunov time (a characteristic time
of the formal instability) and the event time, i.e.,  the time after which a
physically significant change of the planetary configuration happens \cite[see,
e.g.,][]{Lecar2001,Michtchenko2001}.  Still,  the chaotic motions may easily
destabilize the planetary configuration over a short-time scale related to the
relevant low-order MMRs. It can be explained for the 2-planet close to coplanar
systems. The recent secular theories of  \cite{Michtchenko2004,Lee2003} predict
that the main sources of short-time instabilities are related to low-order MMRs
or proximity  of the system to collision zones. Outside the resonances and far
from  the collision zones, the planetary system is generically stable, even in
the range of large eccentricities. These works generalize the Laplace-Lagrange
secular theory \citep{Dermott1999}. For the $N$-planet system,
\cite{Pauwels1983} derived a  similar conclusion, nevertheless  it is formally
restricted to the range of small eccentricities. In the neighborhood of the
collision zone, the MMRs overlap and that leads to the origin of a region of
global instability. The fitting process should certainly eliminate initial
conditions in such zones and, in general, strongly chaotic motions related to
unstable regions of the MMRs.   Fortunately, these can be detected numerically
thank to efficient fast indicators over characteristic event time-scale which is
counted roughly in $10^4$--$10^5$ of the longest orbital periods.

In this paper, we reanalyze the RV data for \stara{} \citep{Correia2005}, 
\starb{} \citep{Naef2004}, \starc{} and \stard{} \citep{Vogt2005} using GAMP.  A
common feature of these systems is that the available measurements cover only
partially or a small number of the longest orbital periods.  The planetary
systems  reside in the zones spanned by strong low-order MMRs. This work
extends  the results of our recent papers devoted to $\mu$~Arae 
\citep{Gozdziewski2003b,Gozdziewski2003e}, HD~82943 and HD~123811
\citep{Gozdziewski2005d}. The studied systems are selected as representative
cases found among the detected multi-planet configurations.

\section{The numerical setup and the fitting method}
%
In order to incorporate the theoretical ideas described in the previous section
we  employ a few numerical tools merged in a self-consistent manner. To
efficiently explore the phase-space (whose structure is understood in terms of
the KAM theorem), we use the Genetic Algorithms scheme (GAs) implemented by
\cite{Charbonneau1995}. The GAs makes it possible, in principle, to find the
global minimum of $\Chi$. In the GAMP code, the solutions to which the GAs
converged are finally refined by a very accurate non-gradient simplex scheme by
Nelder and Mead \citep{Press1992}.  The fractional convergence tolerance to be
achieved in the simplex code is set in the range $10^{-4}$--$10^{-6}$
(typically, the lower accuracy is forced in time consuming GAMP tests). The
simplex refinement in the CPU expensive GAMP code reduces the CPU usage
dramatically by factor of tens\footnote{The fitting code may be then called
GAMPS  (Genetic Algorithm with Megno Penalty and Simplex).}. Yet a single 
particular initial condition may be fine-tuned to match required (or possible to
obtain) accuracy. 

The reflex motion of a star is described with the self-consistent Newtonian
$N$-body  model \citep{Laughlin2001}. The character of  planetary motions is
determined in terms of the  Lyapunov characteristic exponent  which is expressed
by the MEGNO indicator \citep[]{Cincotta2003}.  Thank to an excellent
sensitivity of this indicator to chaotic motions (in particular, accompanying
close encounters), very short integration times are sufficient to remove the
most unstable initial conditions. Typically, the computations rely on $\sim
10^3$--$10^4$ orbital periods of an outermost body. This is not long enough to
eliminate all chaotic motions, but we are left (and in fact it is a desired
feature of the method) with regular or mildly chaotic configurations typically
located  on the borders of stable zones. Let us note that the GAMP code may
basically use any arbitrary stability criterion. In this sense, the method is
quite general. Nevertheless, the definition of the KAM-stability which is
directly  related to the Lyapunov exponent seems to be the most natural and well
justified by the theoretical considerations.

The dynamical neighborhood of a best-fit solution is examined by other fast
indicators. The first indicator derived on the basis of the Fast  Fourier
Transform (FFT) is called the spectral method \cite[SM;][]{Michtchenko2001}.   A
refined and more complex method of this type is the Frequency Analysis by
\cite{Laskar1993}. In our work SM is employed to resolve the structure of the
spectral signal produced by short-term dynamics (i.e., the proper mean motion as
one of fundamental frequencies). After many comparative tests we found that both
MEGNO and SM are similarly sensitive to the chaotic diffusion generated by the
MMRs in  systems with Jupiter-like planets. To detect it, the required
integration time is relatively very short, typically about $10^4$ periods of the
outermost planet.  Under some conditions, SM is even more efficient than MEGNO
because one avoids integrating complex variational equations.  It also provides
a straightforward identification of the MMRs.  The SM is used for computations
of dynamical maps in 2D planes of  selected osculating  elements.  Yet another
fast indicator which helps us to detect physically  significant changes of the
orbital configurations is the $\max e$ indicator (the maximal eccentricity
attained by the orbit of the investigated planet during a prescribed integration
time). We use all three fast indicators as they complement each other. This
makes it possible to examine the dynamical properties of the best-fit solutions
through different characteristics of the  dynamics: the  maximal Lyapunov
exponent, the variation of the fundamental frequencies  and the geometrical
evolution of orbits. 

\section{A planetary system in an exact MMR (HD~202206)}
%
The 2-planet system around \stara{} was discovered by the Geneva Planet Search
Team \citep{Correia2005}. In this system a massive  Jupiter-like planet or a
brown dwarf is accompanied by  a smaller Jovian body in a more distant orbit.
The analysis conducted by \cite{Correia2005} revealed that both planets are
likely involved in a 5:1~MMR. They found that both, the best 2-Keplerian and
$N$-body  solutions are very unstable, and lead to a disintegration of the
system during a  few thousand of years. In order to find a dynamically stable
solution,   the stability map (in terms of the diffusion rate of the proper mean
motion)  in the neighborhood of the best Newtonian fit has been computed. Next,
by  a rather arbitrary post-fit adjusting of the orbital parameters, a stable 
configuration has been selected.  Certainly, we should not expect that such 
changes of the initial condition will provide a statistically optimal result. 

The RV data of \stara{} could be used to test our approach and the quality of
the  dynamical analysis by the discovery team. Unfortunately, the full set of
RVs has  not been published. To overcome this problem, we scanned the relevant
figure from \cite{Correia2005}. A large fraction of the measurements has been
published in an  earlier paper \citep{Udry2002}. This set comprises 95
measurements. Some of them (apparently the most uncertain ones) have been
removed and do not appear in the  data set used by \cite{Correia2005}
(105~measurements).  By comparing the common  part of the data sets, we can
precisely  estimate the quality of the scanned ''measurements''. We found that
the standard deviation of the differences between the relevant data points in
scanned and published measurements is less than 1~m/s in radial velocities and
$\sim 0.3$~d in the moments of observations. These "errors'' are very small
compared to large variations of the RV signal ($\sim 1200$~m/s) and long orbital
periods. Indeed, using the synthetic data we can recover the solutions by
\cite{Correia2005}. 

The results of the GAMP analysis are illustrated in Fig.~\ref{fig:fig1}.  The
code run a hundred times, and we collected the solutions to which the procedure
converged. The parameters of these solutions are illustrated by projections onto
the representative planes of the osculating elements at the initial epoch of the
first observation  (note that to directly compare the results obtained by the
discovery team and by us, the RV measurements are not rescaled by the stellar
jitter). In quite an extensive search we looked only for coplanar
configurations.  In Fig.~\ref{fig:fig1} we mark only {\em stable} solutions with
$\Chi=1.65$ (small filled circles). That value of $\Chi$ is comparable with
$\Chi$ of the best stable fit S5 from \citep{Correia2005}. The application of
GAMP makes it possible to find a better solution with $\Chi=1.52$ and an rms  of
$\sim 10$~m/s. The stability analysis of this fit is illustrated in the left
panels of  Fig.~\ref{fig:fig2}. The solution can be found close to the border of
a relatively narrow stable island of the 5:1~MMR (the left-upper panel for $\log
SN$). In the same integration we computed the indicator $\max e_{\idm{c}}$ (the
left-bottom panel of Fig.~\ref{fig:fig2}). An almost perfect coincidence of
these two plots is striking. It means that the formally chaotic solutions are
physically unstable in the sense that their configurations disrupt rapidly, at
most during the integration period of $7\cdot 10^5$~yr or  $\sim 2\cdot 10^4
P_{\idm{c}}$. Another conclusion is that one should not skip the stability test
in the fitting procedure, as the procedure will not "see'' the rapidly changing
regions of the permitted (stable) initial conditions. To illustrate this issue,
we computed the dynamical maps for a marginally worse solution (with
$\Chi=1.627$ and an rms $\sim 10.32$~m/s; its orbital parameters are given in
the caption to Fig.~\ref{fig:fig2}). In this  case the MMRs 4:1, 5:1, 6:1
overlap and the resulting stable zones are much wider  than for the formal
best-fit! Let us note that in this solution $a_{\idm{c}}$ is  larger by about
0.1~AU from $a_{\idm{c}}$ of the best-fit solution and can be found  in a small
clump of points in Fig.~\ref{fig:fig1} to the right of the main minimum.  Yet
the close coincidence between the $\log SN$ and $\max e_{\idm{c}}$ maps is still
accurately  preserved. This constitutes an excellent argument for the validity
of the GAMP-like  approach. Without it, stable solutions can basically be found
only by chance.

In another search we assumed that the system is not coplanar. We extended the
model to 14 free parameters including the orbital inclinations and one nodal
longitude. As one would expect, the  inclinations are barely constrained by the
RV data, nevertheless, we found an interesting behavior of the \stara{} system.
We found many solutions whose initial orbits have similar inclinations but the
relative inclination is not small, $r_{\idm{rel}} \sim 94^{\circ}$.
We selected one of such solutions for a closer analysis. Its parameters are
given in the caption to Fig.~\ref{fig:fig3}. The synthetic RV curves for both
solutions (the coplanar and the mutually inclined configurations) can be barely
distinguished one from another (see Fig.~\ref{fig:fig4}). The best-fit
solution with the inclined orbits is also found in the zone of the 5:1~MMR (see
Fig.~\ref{fig:fig3}). Again, the formal stability is closely related to the
geometrical evolution of the orbits.

Remarkably, the orbital evolution in these two cases is qualitatively different.
This is illustrated in Fig.~\ref{fig:fig5}. In the coplanar fit, as one would
expect, the orbital eccentricity of the more massive planet stays close to the 
initial value, while the eccentricity of the outer planet varies with a large 
amplitude according to the conservation of the total angular momentum.  In the
mutually inclined configuration, the orbital evolution is quite unexpected. The
eccentricity of the {\em inner and larger} planet varies with a much larger
amplitude than the eccentricity of the smaller companion. Simultaneously, the
inclination of the companion~c spans almost the whole possible range.  This
example demonstrates a potential  problem with a proper interpretation of the RV
data. Since we do not know  the true initial orbital inclinations, we cannot be
sure about the choice of  the best-fit configuration, and hence the orbital
evolution of the whole system.  In the case of the \stara{} system this issue is
of special importance because the system may be considered as a hierarchical
one: the inner pair is the binary of the Sun-like star and a brown dwarf; the
outer planet is a Jovian  companion. In this sense the \stara{} is a triple
stellar system rather than a "usual" planetary system. In that case the
assumption of a coplanar configuration  may be no longer valid. Additionally,
some recent works indicate the extrasolar planetary systems with mutually
inclined orbits may be quite frequent \citep{Thommes2003,Adams2003}.  
\section{A trend in the RV data (14~Her)}
%
In many observed extrasolar planetary systems, linear trends in the RV data are
present, indicating the existence of more distant companions. The GAMP  may be
very useful to constrain orbital parameters when the RV observations cover
partially  the longest orbital period. A good example is the $\mu$~Arae system
\citep{Jones2002b,McCarthy2004}. We did an extensive analysis of the available
RV measurements of $\mu$~Ara in two earlier  papers
\citep{Gozdziewski2003e,Gozdziewski2005a}. The results of the first work, based
on the RV data covering only a fragment of the orbital period of the outer
planet, remarkably coincide with the outcome of the second analysis. In the
later work, the observations  cover about of $70\%$ of the orbital period of the
putative outermost planet. We found  that the stability constraints help to
remove the artefacts as the extremely large eccentricity of the outer planet and
provide tight bounds on the space of permissible orbital parameters.

The 14~Her system was  announced by M.~Mayor (1998, oral contribution). The
presence  of a Jovian planet in this system was next confirmed by
\cite{Butler2003} and \cite{Naef2004}. In the later work, the discovery team
found that the RV data have a linear slope of $\sim 3.6$~m/s per year. The
single planet Keplerian solution  yields an  rms about of $14$~m/s. Even the
drift is accounted for, the rms of the single planet+drift model leads to an rms
$\sim 11$~m/s, much larger than the mean observational uncertainties
$\sigma_{\idm{m}} \sim 7.2$~m/s. Because the trend is similar to the one
observed in the $\mu$~Ara data, we try to find a better solution with GAMP.

The 14~Her is a quiet star with $\log R'_{\idm{HK}}=-5.07$ \citep{Naef2004} thus
it is reasonable to adopt a rather safe estimate of the stellar jitter
$\sigma{\idm{j}}=4$~m/s \citep{Wright2005}.  Still, the rms  excess the joint
uncertainty $\sigma=(\sigma^2_{\idm{m}}+\sigma^2_{\idm{j}})^{1/2} \sim 8$~m/s by
a few m/s.  Luckily, the RV data of this star are published and publicly
available \citep{Naef2004}. There are 119 known observations. We combined them
with much more accurate 35 measurements (the mean of their $\sigma_{\idm{m}}\sim
3.1$~m/s) by the Carnegie Planet Search Team \citep{Butler2003}, covering the
middle part of the joint observational window. The full set consists of 154
measurements spanning about $3400$~d and corresponding to about $2 P_{\idm{b}}$.
For the mass  of the parent  star we adopted the value  of  $0.9~M_{\sun}$ 
\citep{Naef2004}. 

We assume that the drift and large residual signal is due to a long-period
companion in the system. Using GAMP, we searched for a body in an orbit of
$a_{\idm{c}} \in (4,10)$~AU, trying to verify whether the available data can be
already useful to constrain the orbital parameters of the putative distant
planet. The results of thousands of independent  GAMP runs are illustrated in
Fig.~\ref{fig:fig6}. Only the parameters of the {\em stable} best-fits are
projected onto selected planes of the osculating elements at the epoch of the
first observation, JD~2,449,464.5956. The better quality of the fits, measured
by their $\Chi$, corresponds to larger symbols. The largest circles are for the
best fit solutions (given in Table~1) having $\Chi \sim 1.11$  an rms $\sim
8.5$~m/s. The smallest filled dots are for the fits with $\Chi \sim 1.4$
corresponding to the limit of rms $\sim 11$~m/s.   Curiously, two well defined
local minima with almost the same value of $\Chi$ are present.  The synthetic RV
curves for the best fits are illustrated in Fig.~\ref{fig:fig7} (all the 
available measurements are also marked with  error bars). As we could expect,
both curves cannot be distinguished from each other in the time-range covered by
the data. However,  a clear choice between the curves could be done already at
the time of writing this paper  (note that a vertical line about of
JD~2,453,736.46 is for the end of the year 2005).

The two best fits  correspond to qualitatively different configurations with
$a_{\idm{c}}\sim 5.8$~AU and $a_{\idm{c}}\sim 9$~AU.  They are found in the
proximity of the 3:1~MMR (14~Her$^a$) and 6:1~MMR (14~Her$^b$), respectively
(see the upper-right panel of Fig.~\ref{fig:fig6}).  Simultaneously, the
parameters of the inner planet, as well as the relative phases of the
companions, are already constrained very well. This makes it possible to perform
a representative test of the  system stability. We calculated two maps centered
at $a_{\idm{c}}$ in the best fits, in the $(a_{\idm{c}},e_{\idm{c}})$-plane,
keeping other orbital elements fixed at their best fit values (Table~1). The
results are illustrated in Fig.~\ref{fig:fig8}: the panels in the left column
are for the best fit 14~Her$^a$ (marked by a crossed circle) and panels  in the
right column are for the best fit 14~Her$^b$.  Clearly, the dynamical map of
$\log~SN$ strongly coincides with the physical stability (in terms of $\max
e_{\idm{c}}$) in the region spanned by low-order resonances 3:1, 7:2 and 4:1.
The best fit 14~Her$^a$ is found close to the border of the 3:1~MMR. This is
illustrated in Fig.~\ref{fig:fig9} for an initial condition which is close to
the best one. It shows, in subsequent panels, time-evolution of the
eccentricities, the angle of the secular resonance $\theta$ and one of the
critical arguments of the 3:1~MMR ($\theta_{\idm{31}}=3\lambda_{\idm{c}}-
\lambda_{\idm{b}}-\varpi_{\idm{b}}-\varpi_{\idm{c}}$). A perfect convergence of
$\Ym(t)$ confirms that the configuration is close to a quasi-periodic, ordered
motion. Quite surprisingly, the zone of stable motion extends up to the
proximity of the collision zone which is marked by a smooth line in the maps. We
note also a very sharp border of the stability regions. Outside these zones, the
configurations disrupt catastrophically which is indicated by $e_{\idm{c}}$
increasing to~1 during at most $10^5$~yr (the integration time). Obviously, in
such a case, using the pure $N$-body model of the RV we  would be not taking
care of the sophisticated, discontinuous structure of the phase-space.   

The last conclusion is also valid for the solution  14~Her$^b$ with the more
distant planet, see the two panels in the right column of Fig.~\ref{fig:fig8}.
We notice an excellent coincidence of the distribution of the best fits
(Fig.~\ref{fig:fig6}) with the stable areas  unveiled by the SM in
Fig.~\ref{fig:fig8}. The space between $[7,9]$~AU is spanned by a few low-order
MMRs  (4:1, 5:1, 6:1) of varying width and already overlapping for $e_{\idm{c}}$
less by $\sim 0.2$ than the values determined by the equation of collision line.
The border of stability areas is  sharp. Formally chaotic configurations would
be quickly disintegrated by collisions or ejection of a companion from the
system (see the relevant $\max e_{\idm{c}}$ map in Fig.~\ref{fig:fig8}). 
Curiously, the best fit is located between 11:2 and 6:1~MMRs what reminds us the
HD~12661 planetary system \citep{Fischer2003,Gozdziewski2003d}. 

According to the above investigations,  it remains very likely that the 14~Her
hosts two Jovian planets involved in an low-order MMR.  Let us remark that the
presence of the two well separated local minima of $\Chi$ is not so clear when
both the RV data sets are analyzed separately.  At present, even if more data
for the 14~Her are available, it would be desirable to  search for the best fits
with a  GAMP-like  algorithm.  Contrary to the impression caused by the presence
of the apparent long-term drift in the data, a few recent measurements may be
already  helpful to resolve a plausible orbital configuration of 14~Her system.
Yet the measurements gathered by the two observing teams are in excellent
accord, and the data sets are complementing  each other. 

\section{A multi-planet configuration (HD~37124)}
%
Recently, \cite{Vogt2005} announced a discovery of several  multi-planet
systems.  In particular, the formerly known 2-planet system about HD~37124 is
supposed to harbor one more planet. A hypothesis of the third planet removes a
previously present degeneracy of the 2-planet solution to the RV allowing large
eccentricity of the outer planet and collisional destabilization of the system
\citep{Gozdziewski2003b}.   The discovery team found that the dual-Keplerian
model of the RV reveals two, similarly good, best fit solutions. In the better
one, the planets would revolve in almost circular orbits with periods of about
155~d, 843~d and 2300~d respectively. Curiously, in this solution Keplerian
periods $P_{\idm{d}} \sim 3P{\idm{c}}$ may indicate a proximity of the system to
a low-order resonance.  A possibility of such low-order commensurability warns
that the application of the Keplerian models of the RV is problematic. Indeed,
an inspection of the 3-planet, Keplerian best-fit solutions reveals that they
correspond to strongly chaotic motions and  the system easily disintegrates
through mutual interactions. The 3-Keplerian initial condition found by
\cite{Vogt2005} has fixed $e_{\idm{d}}=0.2$ which is chosen to fulfill the
requirement of dynamical stability.

The HD~37124 is a quiet star, with low activity index $\log R'_{\idm{HK}}=-4.90$
\citep{Vogt2005} so the stellar jitter is likely small. We follow the discovery
team by adopting $\sigma_{\idm{j}}=3.2$~m/s. The instrumental errors have been
rescaled by adding that value of jitter in quadrature to the measurement
uncertainties. Next we reanalyzed the RV data with GAMP by conducting two
searches. Having in mind the results of the Keplerian-based analysis done by the
discovery team,  in the first search  we assumed that  the companions orbits
have moderate eccentricities, in the range of  $[0,0.5]$ and semi-major axes in
safe enough bounds of  $a_{\idm{b}} \in [0.4,0.8]$~AU, $a_{\idm{c}} \in
[1.0,1.8]$~AU, $a_{\idm{d}} \in [2.0,4.0]$~AU. The MEGNO was computed for the
whole system  over $\sim 2\cdot 10^3$ periods of the most outer planet. The
results of that GAMP search are illustrated in Fig.~\ref{fig:fig10}. The
subsequent panels are for the projections of the best-fit parameters onto the
planes of osculating elements at the epoch of the first observation. The quality
of gathered solutions is marked by size of symbols. The smallest  filled circles
are for fits having $\Chi<1.14$, i.e., within the $3\sigma$ confidence interval
of the best-fit solution given in Table~1.  It appears that the found single
minimum of $\Chi$ is quite precisely determined. The elements of the most inner
planets are already very well constrained. For instance, the semi-major axis of
companion~b changes within  only 0.002~AU at the $1\sigma$ confidence interval
of the best fit solution. Obviously, the largest uncertainties are for the most
outer companion~d but even in this case the errors are not large. 

Yet the apparently well constrained minimum of $\Chi$ is localized in a region
of the phase space which has a very complex  dynamical structure. This is
illustrated in  Fig.~\ref{fig:fig11} which is for the dynamical maps in the
$(a_{\idm{d}},e_{\idm{d}})$-plane. These maps are computed for a few initial
conditions chosen from the set of the best fits illustrated in
Fig.~\ref{fig:fig10}.  The  relevant initial conditions are marked in the
dynamical maps by large crossed circles. The left-upper panel of
Fig.~\ref{fig:fig10} is for the best Newtonian  solution obtained {\em without}
the MEGNO penalty.  Mathematically, the fit is the best one as its $\Chi \sim
0.86$ and an rms $\sim 3.11$~m/s, a little better than the best-fit to the
3-Keplerian model quoted by \cite{Vogt2005} yielding $\Chi \sim 0.89$.
Nevertheless, this fit is dynamically unacceptable because it lies very close to
the collision zone of the two outermost orbits which is marked by a smooth
curve. In this area the motions are strongly chaotic and unstable.    Far below
the collision line,  we identify the most relevant MMRs of these planets: 7:4,
5:2, 8:3 and 3:1, at the very edge of the map. In the right-upper panel of
Fig.~\ref{fig:fig10}, we choose a relatively good initial condition with
$\Chi\sim1.25$ and an rms $\sim 4.4$~m/s which is  located in a proximity of the
7:3~MMR. Note a significant change of the shape of the 5:2~MMR as compared with
the previous panel. Some fits at the $1\sigma$---$2\sigma$  confidence levels of
$\Chi$ may fall into the libration zone of this MMR and they have quite large
initial $e_{\idm{d}} \sim 0.3$.   An example is illustrated in
Fig.~\ref{fig:fig13}. The configuration is formally chaotic, but the critical
angles  $\theta=\varpi_{\idm{d}}-\varpi_{\idm{c}}$ and \
$\theta_{52}=5\lambda_{\idm{d}}-2\lambda_{\idm{c}}-3\varpi_{\idm{d}}$  librate
about $180^{\circ}$; the eccentricities do not exhibit any secular changes over
3~Myr integration.  Note that in this case MEGNO stays close to 2 for about
0.3~Myr, the stability criterion used in GAMP was not violated and the weakly
chaotic configuration has been left in the set of acceptable solutions.

The left-bottom panel in Fig.~\ref{fig:fig10} is for the  best  fit with
$\Chi=1.11$ and an rms $\sim 4$~m/s, with small initial eccentricity of the most
outer planet. The last one, right-bottom panel is for the initial condition
which is very close to the {\em stable}  best fit solution whose parameters are
given in Table~1. Its $\Chi\sim1$ and an rms $\sim 3.62$~m/s. It would
correspond to a system locked in  the 11:4 MMR of the most outer planets. In the
last panel,  for a reference, we marked the best-fits within $\Chi<1.01$ roughly
corresponding to  the $1\sigma$ confidence interval of the best fit solution. 
Actually, many fits found in this zone, which are computed with a relatively 
very short integration time of MEGNO, $\sim 2000$~$P_{\idm{c}}$, appear to be
mildly chaotic. Note that some of them, including the best one, are found close
to the border of 5:2~MMR. 

Let us remark that the best-fit initial condition given in Table~1 has been
refined through post-fitting with the MEGNO computed over 12,000 periods of the
outer planet and it yields  a close to quasi-periodic configuration.   We
computed MEGNO for this fit over 3~Myrs, and we found that the indicator very
slowly diverges with a rate corresponding to the Lapunov time of $\sim 10^8$~yr.
Its synthetic RV  curve is shown in Fig.~\ref{fig:fig12}. The initial
eccentricities of the two most inner planets are close to 0, nevertheless, it
does not mean that the planets move on close to circular orbits. In fact, all
eccentricities change with a significant amplitude of $\sim 0.2$ --- 0.25.

In the relevant region of the $(a_{\idm{d}},e_{\idm{d}}$)-plane, the positions
of the MMRs, as well as their widths,  vary in the range of $\sim 0.2$~AU with
respect to $a_{\idm{d}}$ when the initial conditions are changed. The border of 
the zone of global instability is highly irregular but very sharp and, as one
would expect, it can be found in the $\max e$ maps (not shown here). A
conclusion provided by this experiment is  that the structure of the phase space
changes dramatically, even if we choose statistically comparable, relatively
close each to other initial conditions. An inspection of the dynamical maps in
Fig.~\ref{fig:fig11}  reveals that it is hardly  possible to avoid the unstable
areas without explicitly accounting the stability criterion in a self-consistent
manner.  One might think that the $N$-body model does not lead to a significant
improvement of $\Chi$ --- we obtained very similar values of $\Chi \sim
0.96$---$0.98$ to those of the best fits found with the triple-Keplerian model
of the RV. Nevertheless, both the Keplerian and Newtonian  best-fit solutions
obtained without the  stability check yield physically unacceptable, disrupting
configurations. 

The best fits found in the GAMP search  reveal an intriguing state of the
HD~37124 system. It resides in a dynamically very active region of the
phase-space. It remains  possible that the two external planets are close to the
5:2 MMR, similarity to the Jupiter-Saturn case.  We found some acceptable fits
within the libration island of this resonance, however, in such a case the
eccentricities of both the most outer companions would be relatively large $\sim
0.3$ (see Fig.~\ref{fig:fig13}). Some  best-fit configurations are very close to
the 8:3 or 11:4~MMR.  As we demonstrate by the computations illustrated in
Fig.~\ref{fig:fig10}, the parameters' errors bounds are relatively extended and
the proximity of the system to any of these resonances cannot be excluded at
present.  

In the second GAMP search we looked for the best fits assuming that the
semi-major axes are about of  $a_{\idm{b}} \in [0.05,0.3]$~AU, $a_{\idm{c}} \in
[0.3,0.6]$~AU, and $a_{\idm{d}} \in [1.2,1.8]$~AU.  In this way we tried to
verify the Keplerian fits of the second plausible  configuration found by the
discovery team. Their analysis reveals the best fit solution which could be
concurrent to the configuration with the long-period orbit of companion~d but
having a significantly worse $\Chi \sim 1.14$ \citep{Vogt2005}.  The
GAMP-resolved $N$-body solutions are also not so good as for the  previously
analysed configurations. The  best fit found in the GAMP search has $\Chi \sim
1.2$.  In that case the innermost planet would be a hot-Neptune with the mass of
about $0.1$~m$_{\idm{N}}$ and semi-major axis of about $0.1$~AU. In overall,
this fit is even worse than the triple-Keplerian fit found by the discovery
team, with $\Chi\sim1.14$. However, it remains possible (but not very likely)
that we missed a better solution. Yet it could be also  a dynamically derived
argument against the configuration with the hot-Neptune planet.  Another 
argument against such solution is given in the very recent  work by
\cite{Ford2005} who found with Bayesian technique that the short-period orbit
(of $30$~d) is not very credible.

\section{Is the GAMP not always necessary? (HD~108874)}
%
The dual planet system about \stard{} can be close to the 4:1~MMR. That
conclusion follows from the analysis of 2-Keplerian model of the RV by the
discovery team \citep{Vogt2005}. Having in mind the \stara{} system, we suspect
that the GAMP code should help us in better understanding of the system dynamics
than follows from the kinematic approach.  Dynamical simulations which rely on
the dual-Keplerian fit by the discovery team revealed that \stard{} is a
dynamically  active system. Initial conditions derived from the kinematic model
may lead, depending on the initial epoch,  to the destruction of the system in a
time scale of about 0.5~Myr.

According to \cite{Vogt2005} the \stard{}, is an inactive star with $\log
R'_{\idm{HK}} =-5.07$ thus, following the discovery team,  we adopted
$\sigma_{\idm{j}}$ of 3.9~m/s. The results  of the GAMP search are illustrated
in Fig.~\ref{fig:fig14}  which is for the solutions spanning formal
$1\sigma,2\sigma,3\sigma$ confidence intervals of the best fit  (its  parameters
are given in Table~1). In that figure,  the osculating elements of the best fits
gathered by independent runs of the GAMP code are illustrated as projections
onto the planes of orbital elements. This gives us also estimates of the fit
errors.  In the independent runs, the fits converged to the same solution as  
in Table~1. That solution has orbital parameters similar to those found with
2-Keplerian model of the RV, nevertheless, the quality of the fit measured by
$\Chi \sim 0.71$ is slightly better; the double-Kelper model yields $\Chi \sim
0.79$ \citep{Vogt2005}.

It appears that orbital elements of both companions are  already well
constrained through the available RV measurements. An interesting conclusion can
be derived from the inspection of the two first bottom-panels of
Fig.~\ref{fig:fig14}, for $(\omega_{\idm{b}},\omega_{\idm{c}})$- and
$(\lambda_{\idm{b}},\lambda_{\idm{c}})$- planes.  While $\omega_{\idm{b}}$ of
the best fits is spread over  the whole possible range, both planetary
longitudes are very well bounded. It means that the parameters $\omega$ and $M$
(the mean anomaly) may be apparently unconstrained, nevertheless, their sum
gives us a  well fixed orbital phase. Finally, we computed the dynamical maps in
the $(a_{\idm{c}},e_{\idm{c}})$-plane   (the left panel of Fig.~\ref{fig:fig15}
is for $\log SN$ and the right panel of this figure is  for $\max e_{\idm{c}}$).
We notice that the border of formally unstable region  begins well under the
planetary collision line. However, in the libration area of the 4:1~MMR the
stable motions are possible up to $e_{\idm{c}} \sim 0.7$! In the $\log SN$ map
we marked the orbital parameters of the fits within the $1\sigma$ confidence
interval of the  best fit solution. They cover the whole  resonance width  of
about $0.05$~AU. The best fit solution is found close to the separatrix of the
resonance. The synthetic RV curve is shown in Fig.~\ref{fig:fig16}, it perfectly
follows the measurements. Finally, Fig.~\ref{fig:fig17} is for the orbital
evolution of the configuration derived from the best fit (Table~1) and its
stability analysis by MEGNO. MEGNO has been computed for over 10~Myr $(\sim
2\cdot 10^6 P_{\idm{c}})$  and perfectly converges to 2  at this period of time,
so this configuration is strictly quasi-periodic. This is also seen in the time
evolution of the eccentricities --- no secular drifts are present, and their
amplitudes are very small. Actually, the system is locked in the 4:1~MMR as the
one of the critical arguments, 
$\theta_{41}=4\lambda_{\idm{c}}-\lambda_{\idm{b}}-2\varpi_{\idm{c}}-\varpi_{\idm{b}}$
librates about $0^{\circ}$. 

Our conclusions are somehow against the results of \cite{Vogt2005} who concluded
that the system can be currently described by a large number of dynamically
distinct configurations.  Curiously, the best fit found with GAMP is almost the
same as the one derived  without the penalty term and only slightly better
($\sim 0.5$~m/s in the rms) than the one obtained with the 2-Kepler
parameterization. The results of stability analysis we did for the best fit
(Fig.~\ref{fig:fig15})  suggest that in the case of \stard{} the use of GAMP is
not so critical for obtaining stable solutions as we found for the other systems
analyzed in this work.  That is likely due to well constrained  orbital
parameters of the best-fit or a specific shape of the resonance. In fact, its
width is comparable with the fit errors. Still, without explicit computations,
we could not be sure in which region of the phase space the best fit is
localized and how this region looks like.

\section{Conclusions}
%

With the application of GAMP  we found a clear indication of a new, second
planetary companion in the 14~Her system. Remarkably, the data permit two
distinct solutions corresponding to the low-order  mean motion resonances 3:1 or
6:1. A few recent observations about the date of writing this paper could be
very useful to resolve the doubt.  We also found that the two most outer 
planets in the \starc{} system may be close to 5:2 MMR, thus being remarkably 
similar to the Jupiter-Saturn pair. GAMP helped us to found stable
configurations of the \stard{} system and the results support the hypothesis
that the system is locked in an exact 4:1~MMR.

We have shown in this paper that the GAMP performs very well. Indeed,  the idea
has a solid theoretical background.  Applying the obvious requirement of the
dynamical stability, we should  eliminate the initial conditions which lead to a
quick destruction of a planetary configuration. A delicate matter is the
question of how to understand (and measure) the stability. In this paper we
prefer the formal definition provided by the KAM-theorem.  Essentially,  the
dynamics of a planetary system has two time-scales related to the fast orbital
motions and their resonances (MMR's)  and much slower precession of
instantaneous orbits (secular dynamics).   Analyzing the relatively small sets
of the RV measurements, and due to narrow observational windows, we are
naturally limited to the short-time scale.  The recent secular theories by 
\cite{Michtchenko2004,Lee2003} for 2-planet systems and the results of
\cite{Pauwels1983} for a general $N$-planet system in the regime of moderate
eccentricities are very useful to predict the generic features of such systems.
They are generically  stable  under  the condition that planets are not involved
in strongly chaotic motions (usually related to low-order MMRs)  or their orbits
stay far from collision zones. Our line of reasoning is that, at least in the
first approximation, we should eliminate initial conditions corresponding to
such unstable behaviors. It is possible thank to computationally efficient fast
indicators. Yet, according to the KAM theorem, the search for the best fit
solutions is conducted  in a highly noncontinuous parameter space. A cure for
this problem is an application of non-gradient Genetic Algorithms which have
features ideally suited to our purposes. The GAs need only "to know'' the $\Chi$
function and efficiently explore the phase space. To eliminate the unstable
solutions we add a penalty term to the formal $\Chi$ of potential solutions. Let
us underline  that such penalty term may be arbitrary, so in fact we may use
virtually any criterion of stability. Still, one should be aware that the
GAMP-like code is CPU-expensive. For instance, the GAMP calculations typically
occupy through several days a 16-processor AMD/Opteron 2Ghz cluster for every
system studied in this paper. Nevertheless,  the method may be optimized in many
ways. 

The multi-planet configurations analyzed in this paper  are representative cases
in which we may benefit from the application of GAMP-like code.  Frequently, the
RV data span a short time with respect to the longest orbital periods and then
pure Keplerian, or even $N$-body Newtonian, models of the reflex motion of the
parent star yield physically unacceptable configurations which disrupt during
thousands of years. That obviously contradicts the Copernican Principle. A good
example of such situation provides the  $\mu$~Are case \citep{Jones2002b,
McCarthy2004,Gozdziewski2005a} or the 14~Her system~\citep{Butler2003,Naef2004}
analyzed in this paper. In both instances, the RV data indicate  linear trends
over the RV signal of a single planet configuration.  In such instances, the
GAMP-like code makes it possible to limit significantly the otherwise
unconstrained parameters of a putative long-period companions. 

The GAMP-like algorithm is especially well suited for the analysis of RV data of
stars hosting multi-planet systems with Jovian planets likely involved in
strong, low-order MMRs. Such systems are naturally favored by the Doppler
technique because of relatively short observational windows. We have illustrated
the efficiency of the method by analyzing the measurements of 
HD~202206~\citep{Correia2005}, HD~37124 and HD~108874~\citep{Vogt2005}, and
also  HD~128311 \citep{Vogt2005} HD~82943~\citep{Mayor2004} studied in our other
recent paper \citep{Gozdziewski2005d}.  In all these cases, the stability zones
are very sharp and the formal (KAM-like) and astronomical notions of stability
are strictly related to each other. Then, it is essential to use the stability
criterion as an internal part of the fitting algorithm. The GAMP-like code makes
it possible to find the statistically optimal, stable solutions. The stability
analysis is also greatly simplified.  Certainly, the dynamical  analysis of
other resonant systems may also  benefit from the application of this numerical
tool.

According to \cite{Marcy2005a}, nearly all giant planets orbiting within 2~AU of
all FGK stars within 30~pc have now been discovered. The observational windows
of the extrasolar searches are constantly widening.  The orbital periods of
newly revealed, putative planets  become still longer and longer. The full
coverage of their periods by observations has already become a matter of many
years. In this context, the GAMP analysis may be useful to conduct early
detection of long-period planetary companions and to plan the optimal
observational strategy. 

\section{Acknowledgments}
%
We kindly thank Zbroja for the correction of the manuscript.
This work is supported by the Polish 
Ministry of Science and Information Society Technologies, Grant No.~1P03D~021~29.
M.K. is also supported by NASA through grant NNG04GM62G.
K.G. is alco supported by N. Copernicus Univ. through grant 427A.
\bibliographystyle{apj}
%

%
%
%
\bibliography{ms}
%
%
%
{
\hspace*{8cm}
\begin{table}
\caption{\normalsize
Osculating, astrocentric elements of the best fits found in this paper which are
given at the epoch of the relevant first observation. All systems are assumed to
be coplanar and edge-on. Formal estimates of the uncertainties may by derived
from the distributions of best fits which  are illustrated in subsequent figures
in this work. See the text for more details.
}
\smallskip
\normalsize
\begin{tabular}{lccccccccccc}
\hline
&\multicolumn{2}{c}{      HD~202206      } &
 \multicolumn{2}{c}{      14~Her$^a$      }&
 \multicolumn{2}{c}{      14~Her$^b$      } &
 \multicolumn{3}{c}{      HD~37124      } &
 \multicolumn{2}{c}{      HD~108874      } 
\\
Parameter \hspace{1em}  
& \ \  {\bf b} \ \  & \ \ {\bf c} \ \ 
& \ \  {\bf b} \ \  & \ \ {\bf c} \ \
& \ \  {\bf b} \ \  & \ \ {\bf c} \ \   
& \ \  {\bf b} \ \  & \ \ {\bf c} & \ \ {\bf d} \ \  
& \ \  {\bf b} \ \  & \ \ {\bf c} \ \   
\\
\hline
$m \sin i $[m$_{\idm{J}}$] \dotfill 
				& 17.624  &  2.421      
				&  4.485  &  2.086 
				&  4.533  &  6.289       
				&  0.614  &  0.572  & 0.612 
				&  1.358  &  1.008    
\\
$a$ [AU] \dotfill 		&  0.831  & 2.701    
				&  2.727  & 5.810
				&  2.730  & 8.911     
				&  0.519  & 1.630   & 3.070
				&  1.051  & 2.658     
\\        
$e$ \dotfill     			&  0.433  & 0.255  
				&  0.361  & 0.004   
				&  0.357  & 0.101            
				&  0.041  & 0.006   & 0.206
				&  0.068  & 0.252     
\\
$\omega$ [deg]\dotfill 		& 161.41  &  92.73 
				&  22.98  & 197.17
				&  22.88  &  62.97         
				& 329.56  &  284.63 & 95.23
				&  255.76 &  16.65    
\\
$M$ [deg] \dotfill 	&  353.44 &  65.76
				&  322.94 &  17.68
				&  323.78 & 227.45
				&  250.13 & 288.68       & 113.35
				&  13.26  & 32.08     
\\
$\Chi$  \dotfill 		& \multicolumn{2}{c}{ 1.53  } & 
		 		  \multicolumn{2}{c}{ 1.11  } &
		 		  \multicolumn{2}{c}{ 1.11  } &
		 		  \multicolumn{3}{c}{ 0.98  } &
		 		  \multicolumn{2}{c}{ 0.71  } 
\\
rms [m/s] \dotfill 	& 	  \multicolumn{2}{c}{ 9.97  } & 
				  \multicolumn{2}{c}{ 8.53  } &
				  \multicolumn{2}{c}{ 8.51  } &
				  \multicolumn{3}{c}{ 3.53  } &
		 		  \multicolumn{2}{c}{ 3.30  }   
\\
$V_0$ [m/s] \dotfill 	& 	  \multicolumn{2}{c}{ -1.36 } &
				  \multicolumn{2}{c}{-14.81 } &
				  \multicolumn{2}{c}{-55.65 } & 				  
				  \multicolumn{3}{c}{ 7.92  } &
		 		  \multicolumn{2}{c}{ 17.28 } 
\\
$V_1$ [m/s] \dotfill 	& 	  \multicolumn{2}{c}{       } &
				  \multicolumn{2}{c}{-49.76 } &
				  \multicolumn{2}{c}{-90.48 } & 				  
				  \multicolumn{3}{c}{       } &
		 		  \multicolumn{2}{c}{       } 
\\
$p$ 	 		& 	  \multicolumn{2}{c}{ 11    } &
				  \multicolumn{2}{c}{ 12    } &
				  \multicolumn{2}{c}{ 12    } & 				  
				  \multicolumn{3}{c}{ 16    } &
		 		  \multicolumn{2}{c}{ 11    } 
\\
$M_{\star}$~[$M_{\sun}$]\dotfill& \multicolumn{2}{c}{ 1.15  } &
				  \multicolumn{2}{c}{ 0.90  } &
				  \multicolumn{2}{c}{ 0.90  } & 				  
				  \multicolumn{3}{c}{ 0.78  } &
		 		  \multicolumn{2}{c}{ 0.99  } 
\\
$\sigma_{\idm{j}}$\dotfill& \multicolumn{2}{c}{   } &
				  \multicolumn{2}{c}{ 4.0  } &
				  \multicolumn{2}{c}{ 4.0  } & 				  
				  \multicolumn{3}{c}{ 3.2  } &
		 		  \multicolumn{2}{c}{ 3.9  } 
\\
\hline
\end{tabular}
\label{tab:tab2}
\end{table}
}

%
%

\figcaption[f1.eps]{\normalsize
The best fits obtained by the GAMP algorithm for the RV data published 
graphically in \cite{Correia2005} for \stara. The coplanar system  is assumed.
Parameters of the fit are projected onto the planes of osculating orbital
elements at the epoch of the first observation,  JD~2,451,402.8027. The smallest
filled circles are for stable solutions  with $\Chi<1.65$ and an rms $\sim
11$~m/s.  Bigger open circles are for $\Chi<1.55$  and $\Chi< 1.6$ ( the formal
$1\sigma$ confidence interval of the best-fit solution is $\Chi=1.53)$.  The
largest circles are for the solutions with $\Chi< 1.52$, marginally larger than
$\Chi=1.519$ of the best-fit  initial condition.  
}

\figcaption[f2a.eps,f2b.eps,f2c.eps,f2d.eps]{\normalsize
The panels in the left column are for dynamical maps in the
$(a_{\idm{c}},e_{\idm{c}})$-plane in terms of the Spectral Number, $\log SN$ and
$\max e$ for  putative 5:1 MMR in a coplanar \stara{} system (see Table~1).
Colors used in the $\log SN$ map classify the orbits --- black indicates
quasi-periodic, regular configurations while yellow strongly chaotic systems. A
crossed circle marks the best-fit configuration.  The right column is for a
little bit worse initial condition with $\Chi=1.62$ and an rms $\sim 10.32$~m/s;
the osculating  elements at the epoch of the first observation are 
$(m~\mbox{[m$_\idm{{J}}$]},a~\mbox{[AU]},e,
,\omega~\mbox{[deg]},M~\mbox{[deg]})$: (17.589, 0.831, 0.435, 161.118, 353.944)
for planet~b and (2.247, 2.835, 0.220, 159.848, 1.247) for planet~c;
$V_0=-0.47$~m/s.  The resolution of the maps is $600\times120$ data points.
Integrations are for $2\cdot 10^4$ periods of the outer planet ($\sim 7\cdot
10^4$~yr).  The islands of the relevant MMRs are labeled.
}   

\figcaption[f3a.eps,f3b.eps]{\normalsize
Dynamical maps in the $(a_{\idm{c}},e_{\idm{c}})$-plane in terms of the Spectral
Number, $\log SN$ and $\max e$ for the best-fit with mutually inclined orbits in
the \stara{} system. Colors used in the $\log SN$ map classify the orbits ---
black indicates quasi-periodic, regular configurations while yellow strongly
chaotic systems. A crossed circle denotes the best-fit configuration.  The
initial condition yields $\Chi=1.59$,  an rms $\sim 9.97$~m/s (the number of 
fit parameters is 14). The osculating  elements  at the epoch of the first
observation are  
$(m~\mbox{[m$_\idm{{J}}$]},a~\mbox{[AU]},e,i~\mbox{[deg]},
\Omega~\mbox{[deg]},\omega~\mbox{[deg]},M~\mbox{[deg]})$:
(17.723, 0.831,  0.435, 83.625, 265.307, 161.040, 353.921) for planet~b 
and
(2.348,  2.736, 0.178, 82.372, 0.0, 127.813, 40.962) for  planet~c; 
$V_0=-1.77$~m/s.  
The resolution of the maps is $600\times120$ data points. Integrations are for
$2\cdot 10^4$ periods of the outer planet ($\sim 7\cdot 10^4$~yr).
}
   
\figcaption[f4.eps]{\normalsize
The synthetic RV curves for the \stara{} system. The thin line is for a stable
($N$-body) solution corresponding to a 5:1~MMR in the coplanar system, the thick
line is for a 5:1~MMR in the configuration with  mutually inclined orbits.
Circles are for the RV measurements published graphically in
\citep{Correia2005}. 
}

\figcaption[f5.eps]{\normalsize
Orbital evolution of the \stara{} configurations corresponding to the best fit
coplanar solution (the left column, the elements are given in Table~1) and for
the system with mutually inclined orbits (the right column, see the caption to
Fig.~\ref{fig:fig3} for the osculating elements of this configuration).
}

\figcaption[f6.eps]{\normalsize
The best fits obtained with  GAMP for the RV data published in
\citep{Butler2003} and \citep{Naef2004} for 14~Her. The coplanar system  is
assumed. Parameters of the fit are projected onto the planes of osculating
orbital elements at the epoch of the first observation, JD~2,449,464.5956.  
The smallest, filled circles are for stable  solutions  with $\Chi<1.4$ and an
rms $\sim 11$~m/s.  Bigger open circles are for  $\Chi=1.146$, $\Chi<1.129$,
$\Chi< 1.117$ corresponding  to $3\sigma$, $2\sigma$ and $1\sigma$ confidence
intervals of the best-fit solution, respectively.  The largest circles are for
the solutions with $\Chi<1.111$, marginally larger than $\Chi=1.109$ of the two
best-fits given Table~1. A curve in the $(a_{\idm{c}},e_{\idm{c}})$-plane is for
the planetary collision line. It is determined from the relation 
$a_{\idm{b}}(1+e_{\idm{b}})=a_{\idm{c}}(1-e_{\idm{c}})$  with
$a_{\idm{b}},e_{\idm{b}}$  fixed at their best-fit values. The nominal 
positions of the most  relevant MMR inferred from the Kepler law are also marked
by dashed lines and labeled.   
}

\figcaption[f7.eps]{\normalsize
The synthetic RV curves for the two best fit solutions (see Table~1) to the RV
data of \starb{}. The thick line is for 14~Her$^a$ (a proximity to the 3:1~MMR)
and the thin line is for 14~Her$^b$ (about the 6:1~MMR). Data points are plotted
with error bars indicating the joint RV error (stemming from the errors of
measurements and stellar jitter of 4~m/s added in quadrature). The vertical line
about of JD~2,453,736 is for the  calendar date 12/31/2005. 
}

\figcaption[f8a.eps,f8b.eps,f8c.eps,f8d.eps]{\normalsize
The stability maps in the $(a_{\idm{c}},e_{\idm{c}})$-plane in terms of the
Spectral Number, $\log SN$ and  $\max e$ for the best-fit solution to the RV
data of \starb{} system. Colors used in the $\log SN$ map classify the orbits
--- black indicates quasi-periodic, regular configurations while yellow strongly
chaotic systems. The crossed circles mark the best-fit configurations.  The left
column is for 14~Her$^a$ fit, the right column is  for  14~Her$^b$ fit (see
Table~1). The relevant MMRs are labeled. A collision line  according to  the
formulae given in a caption to Fig.~\ref{fig:fig9}, for fixed best fit elements
of the inner planet, is also marked.  The resolution of the maps is
$600\times120$ data points. Integrations are for $2\cdot 10^4$ periods of the
outer planet ($\sim 7\cdot 10^4$)~yr.
}

\figcaption[f9.eps]{\normalsize
The orbital evolution of a configuration close to the best-fit solution of
14~Her$^a$ (see Table~1). The osculating elements for the epoch
of fits observations is terms of
$(m~\mbox{[m$_\idm{{J}}$]},a~\mbox{[AU]},e,\omega~\mbox{[deg]},M~\mbox{[deg]})$
are 
(4.478, 2.726, 0.363, 23.28, 322.706) 
for planet~b and 
(1.945, 5.628, 0.0028, 192.98, 15.33) 
for the planet~c;
$V_0=-13.38$, $V_1=-48.18$~m/s, $\Chi=1.111$, an rms =$8.53$~m/s. 
Subsequent panels are for the eccentricities, the
angle $\theta$ of the secular alignment of the apsides, the MEGNO, $\Ym$,
indicating a quasi-regular configuration and the critical argument of the
3:1~MMR, respectively. 
}

\figcaption[f10.eps]{\normalsize
The best fits obtained with the GAMP for the RV data published in
\citep{Vogt2005} of \starc. The coplanar system  is assumed. Parameters of the
fit are projected onto the planes of osculating orbital elements at the epoch of
first observation, JD~2,450,420.047. 
The smallest filled dots are for GAMP solutions  with $\Chi<1.14$ and an rms 
$\sim 4.1$~m/s. Bigger open circles are for  $\Chi=1.04$, and $\Chi=1.01$
($2\sigma$ and $\sim 1\sigma$ confidence intervals of the best-fit solution,
respectively).  The largest circles are for the solutions with $\Chi<0.961$,
marginally larger than $\Chi=9.56$ of the best-fit found in the whole search. In
the top-right panel, a curve in the $(a_{\idm{d}},e_{\idm{d}})$-plane is for the
planetary collision line for the most outer companions. It is determined from
the relation  $a_{\idm{c}}(1+e_{\idm{c}})=a_{\idm{d}}(1-e_{\idm{d}})$  with
$a_{\idm{c}},e_{\idm{c}}$  fixed at their best-fit values. The nominal 
positions of the most  relevant MMRs inferred from the Kepler law are also
marked by dashed lines and labeled.   
}

\figcaption[f11a .eps,f11b.eps,f11c.eps,f11d.eps]{\normalsize
The stability maps in the $(a_{\idm{d}},e_{\idm{d}})$-plane in terms of the
Spectral Number, $\log SN$ for the best-fit solutions to the RV signal of
\starc{}. Colors used in the $\log SN$ map classify the orbits --- black
indicates quasi-periodic, regular configurations while yellow strongly chaotic
systems. The crossed circles mark the initial conditions used for the
computation of an relevant map.   The initial conditions are given in the terms
of osculating elements at the epoch of the first observation,
$(m~\mbox{[m$_\idm{{J}}$]},a~\mbox{[AU]},e,\omega~\mbox{[deg]},M~\mbox{[deg]})$.
The left-upper panel is for the best $N$-body fit  found {\em without
instability penalty} which 
yields $\Chi= 0.846$ and an rms=3.11~m/s,
(0.614,   0.519,   0.061, 341.18, 238.68),
(0.563,   1.660,   0.070, 163.97,  55.17),
(0.726,   2.973,   0.367, 104.06,  99.19),
for planets b,c,d, respectively and $V_0=7.536$~m/s.
The right-upper panel is for a stable fit with $\Chi=1.25$, an rms=4.43~m/s,
(0.603    0.519,   0.030,   315.80, 264.10),
(0.540,   1.659,   0.061,   179.25,  44.65),
(0.698,   2.915,   0.178,    97.90,  95.57),
for the planets b,c,d, respectively and $V_0=7.26$~m/s.
The left-bottom panel is for a stable solution with $\Chi=1.11$, an rms=4~m/s,
(0.614,   0.519,   0.001,   60.27, 159.73),
(0.640,   1.628,   0.104,  146.01,  69.75),
(0.622,   3.230,   0.005,   10.15, 206.89),
for the planets b,c,d, respectively and $V_0=7.70$~m/s.
The left-bottom panel is for a stable solution in the 
close neighborhood of the 11:4~MMR with $\Chi=1.01$, and an rms=3.63~m/s,
(0.620,   0.519,   0.042,    51.08, 170.00),
(0.592,   1.627,   0.023,    90.86, 122.38),
(0.634,   3.203,   0.202,    95.49, 129.50),
for planets b,c,d, respectively and $V_0=7.78$~m/s.
For a reference, the elements of the best fits with $\Chi<1.01$ obtained  for
the RV data published in \citep{Vogt2005}, at the epoch of first observation,
are marked in the left-bottom panel. Largest circles are for the best stable
solutions with $\Chi<0.98$.  The  smooth curves in the maps mark the collision
line of the two most outer planets. See also Fig.~\ref{fig:fig10}.
The resolution of the maps is $400\times100$ data points. Integrations are for
$6\cdot 10^3$ periods of the most outer  planet ($\sim 3\cdot 10^4)$~yr.
}

\figcaption[f12.eps]{\normalsize
The synthetic RV curves for the best fit solutions  to the RV data of \starc{}
(see Table~1). Data points published in \cite{Vogt2005} are plotted with error
bars indicating the joint RV error (stemming from the measurements and stellar
jitter). The vertical line about of JD~2,453,736 is for the calendar date of
12/31/2005.
}

\figcaption[f13.eps]{\normalsize
The orbital evolution of a configuration close to the best fit  solution of
\starc{} (see Table~1) and corresponding to the libration center of the 5:2~MMR.
Subsequent panels are for the eccentricities, the angle $\theta$ of the secular
alignment of the apsides, the MEGNO, $\Ym$, indicating a quasi-regular
configuration and the critical argument of a 5:2~MMR, respectively.  Parameters
of this fit ($\Chi=1.05$, an rms = 3.82~m/s) in terms of the osculating elements
at the epoch of the first observation,
$(m~\mbox{[m$_\idm{{J}}$]},a~\mbox{[AU]},e,\omega~\mbox{[deg]},M~\mbox{[deg]})$,
are
$(0.633,   0.519,   0.032, 243.56, 334.53)$
for planet~b,
$(0.583,   1.647,   0.015, 304.63, 281.95)$
for planet~c, and
$(0.671,   3.025,   0.269, 127.64,  82.47)$
for planet~d, $V_0=8.61$~m/s.
}

\figcaption[f14.eps]{\normalsize
The best fits obtained with GAMP for the RV data  \citep{Vogt2005} of \stard. In
the model, a coplanar system  is assumed. Parameters of the best fits are
projected onto the planes of osculating orbital elements. The smallest filled
circles are for stable  solutions  with $\Chi<0.91$ corresponding to the
$3\sigma$ confidence interval of the best fit.   Bigger open circles are for
$\Chi<0.82$   and $\Chi<0.76$ (the $2\sigma$  and $1\sigma$ confidence interval
of the best-fit solution given in Table~1, respectively). The largest circles
are for the solutions with $\Chi<0.713$ marginally larger than $\Chi=0.7126$ of
the best-fit initial condition given in Table~1. 
}

\figcaption[f15a.eps,f15b.eps]{\normalsize
The dynamical maps in the $(a_{\idm{c}},e_{\idm{c}})$-plane in terms of the
Spectral Number, $\log SN$ and $\max e$ for the best-fit solution to the
\stard{} RV data.  See Table~1 for the initial condition. Colors used in the
$\log SN$ map classify the orbits --- black indicates quasi-periodic, regular
configurations while yellow strongly chaotic systems.  The resolution of the
maps is $600\times120$ data points. Integrations are for $2\cdot 10^4$ periods
of the outer planet ($\sim 8.6\cdot 10^4$~yr). The parameters of the fits within
$1\sigma$ confidence interval of the best fit  are also marked (see also
Fig.~\ref{fig:fig14}). The crossed circle marks the initial condition used for
computing the maps.
}

\figcaption[f16.eps]{\normalsize
The synthetic RV curve for the best fit solution to the RV data of \stard{} (see
Table~1). Data points published in \cite{Vogt2005} are plotted with error bars
indicating the joint RV error (stemming from the measurements and stellar
jitter). The vertical line about of JD~2,453,736 is for the  calendar date of
12/31/2005. 
}

\figcaption[f17.eps]{\normalsize
The orbital evolution of the best fit configuration of \stard{} (see Table~1).
Subsequent panels are for the eccentricities, the angle $\theta$ of the secular
alignment of the apsides, the MEGNO $\Ym$ and the  critical argument of the
4:1~MMR. MEGNO  is computed over of $2\cdot10^6$ periods of the outermost
planet. A perfect convergence to the value of 2 indicates strictly quasi-regular
configuration. The critical argument of the 4:1~MMR librates about of
$0^{\circ}$ --- the system is locked in an exact 4:1 MMR.
}

\setcounter{figure}{0}
%
%
\begin{figure*}[!h]
   \centering
   \hbox{\includegraphics[]{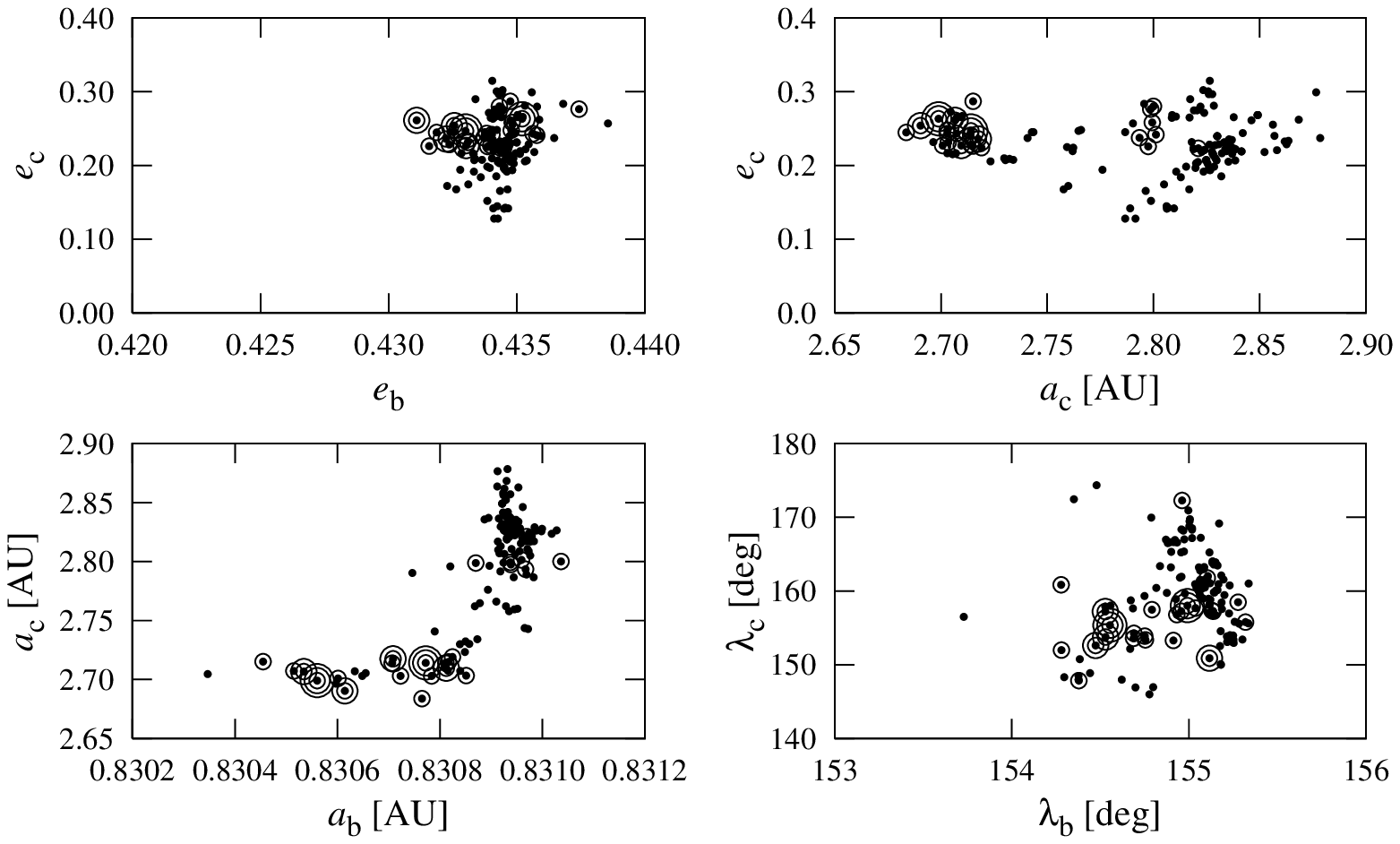}}
   \caption{}
\label{fig:fig1}%
\end{figure*}

%
%
\begin{figure*}
   \centering
   \hbox{\includegraphics[]{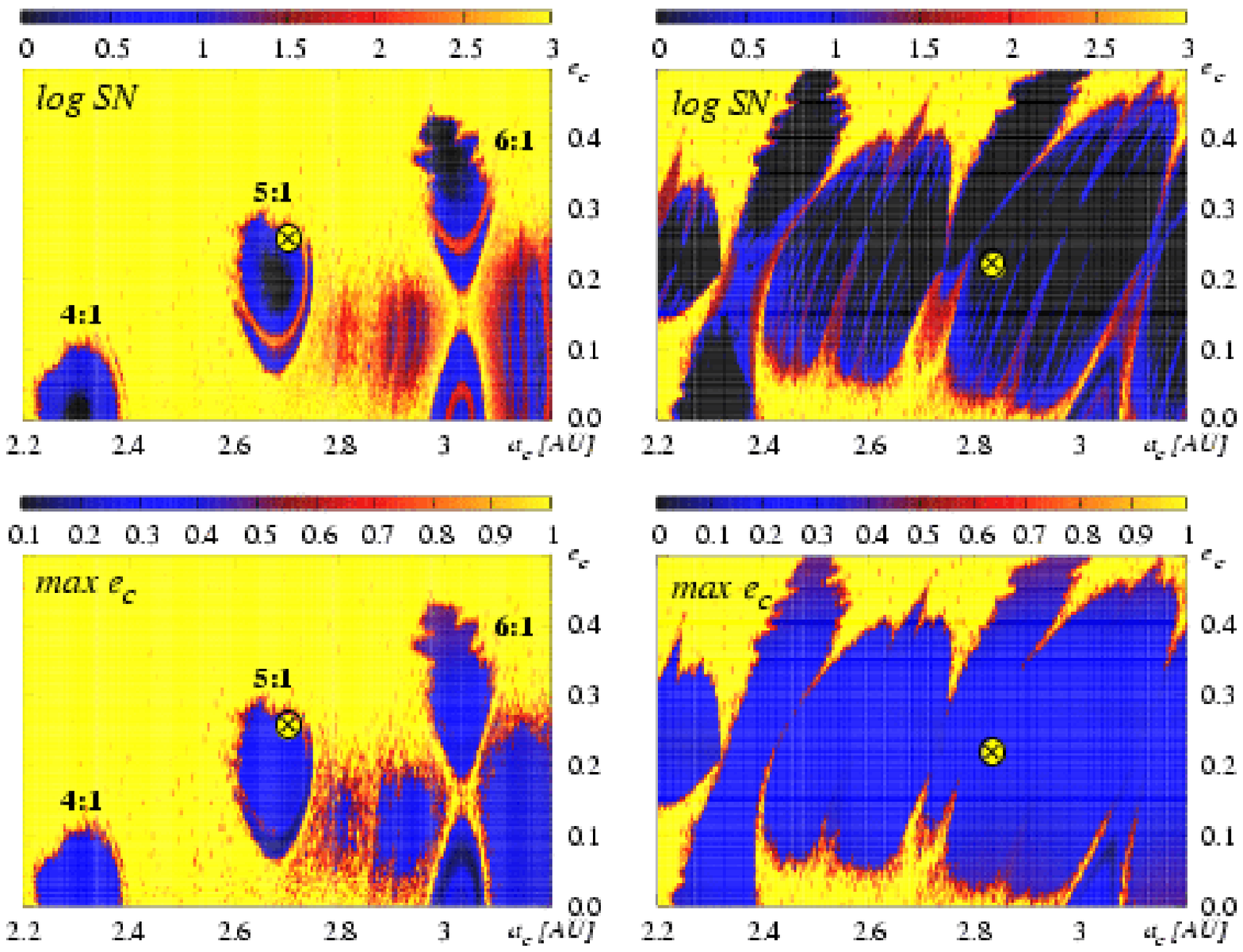}}
   \caption{}
\label{fig:fig2}%
\end{figure*}
%
%
\begin{figure*}
   \centering
   \hbox{\includegraphics[]{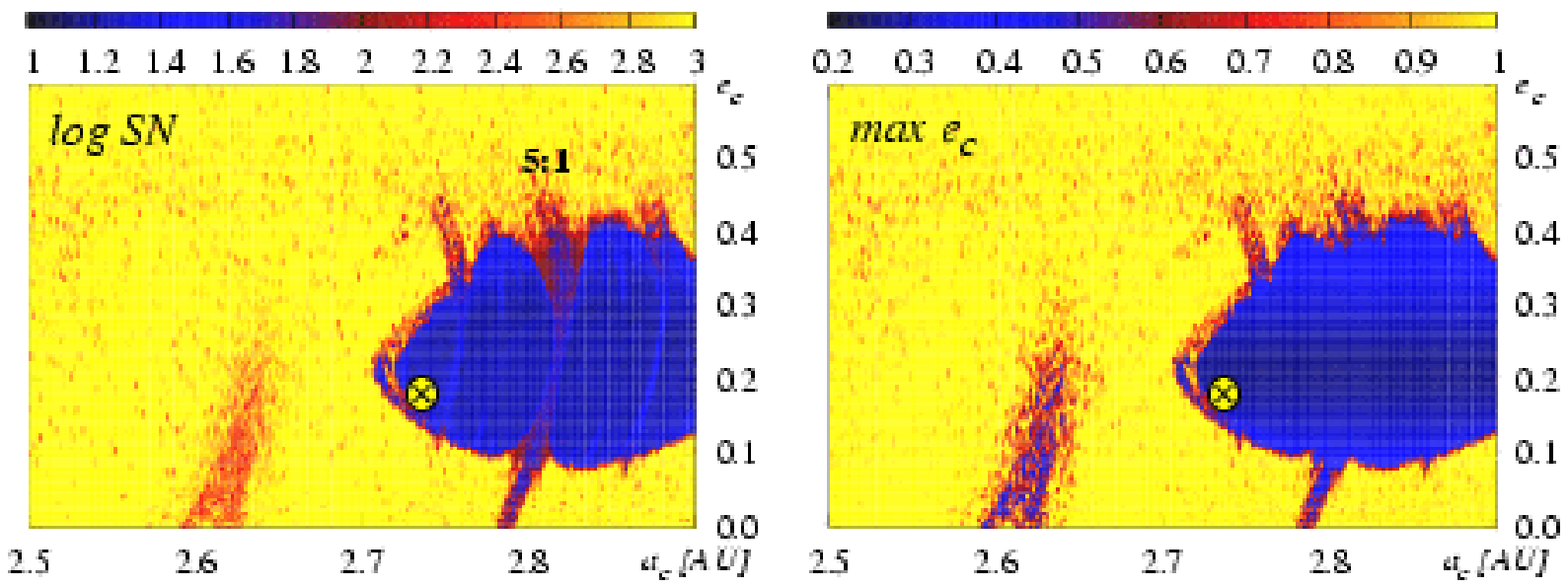}}
   \caption{}
\label{fig:fig3}%
\end{figure*}
%
%
\begin{figure*}
   \centering
   \hbox{\includegraphics[]{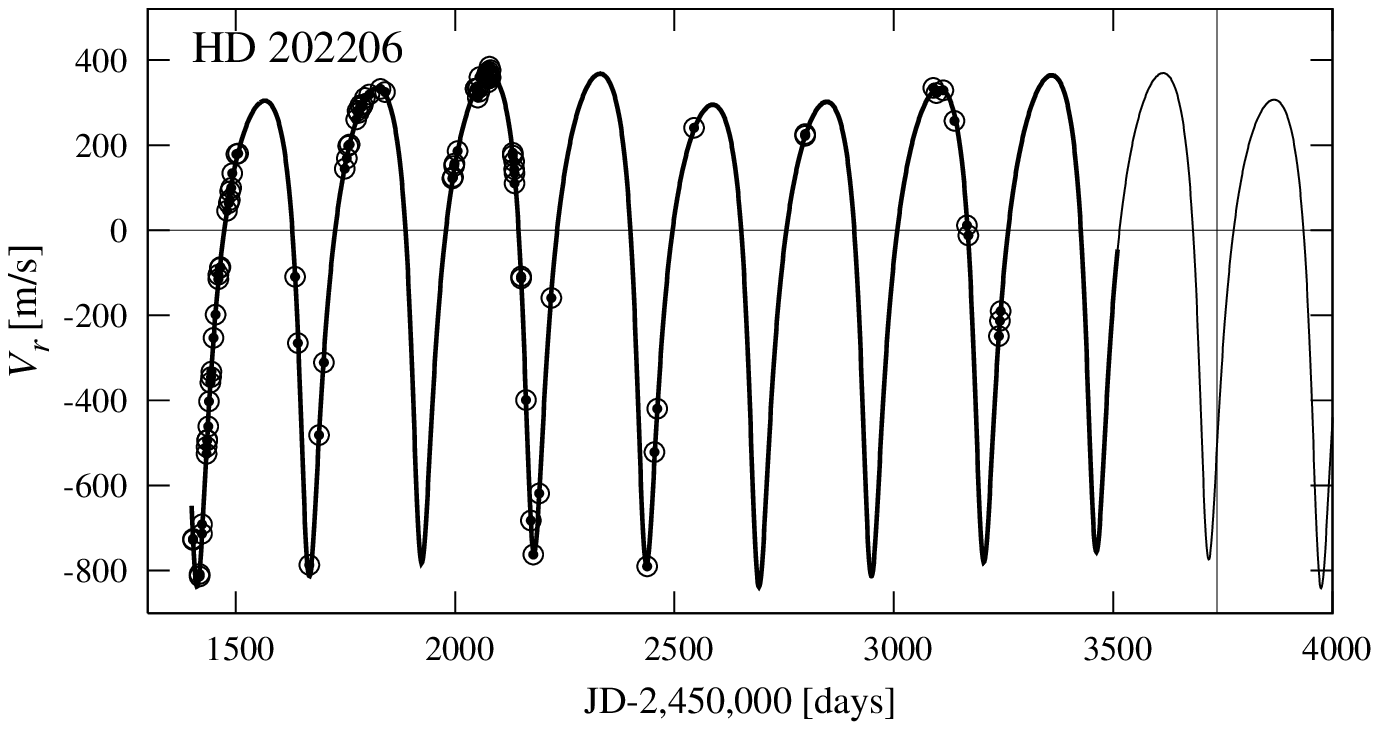}}
   \caption{}
\label{fig:fig4}%
\end{figure*}
%
%
\begin{figure*}
   \centering
   \hbox{\includegraphics[]{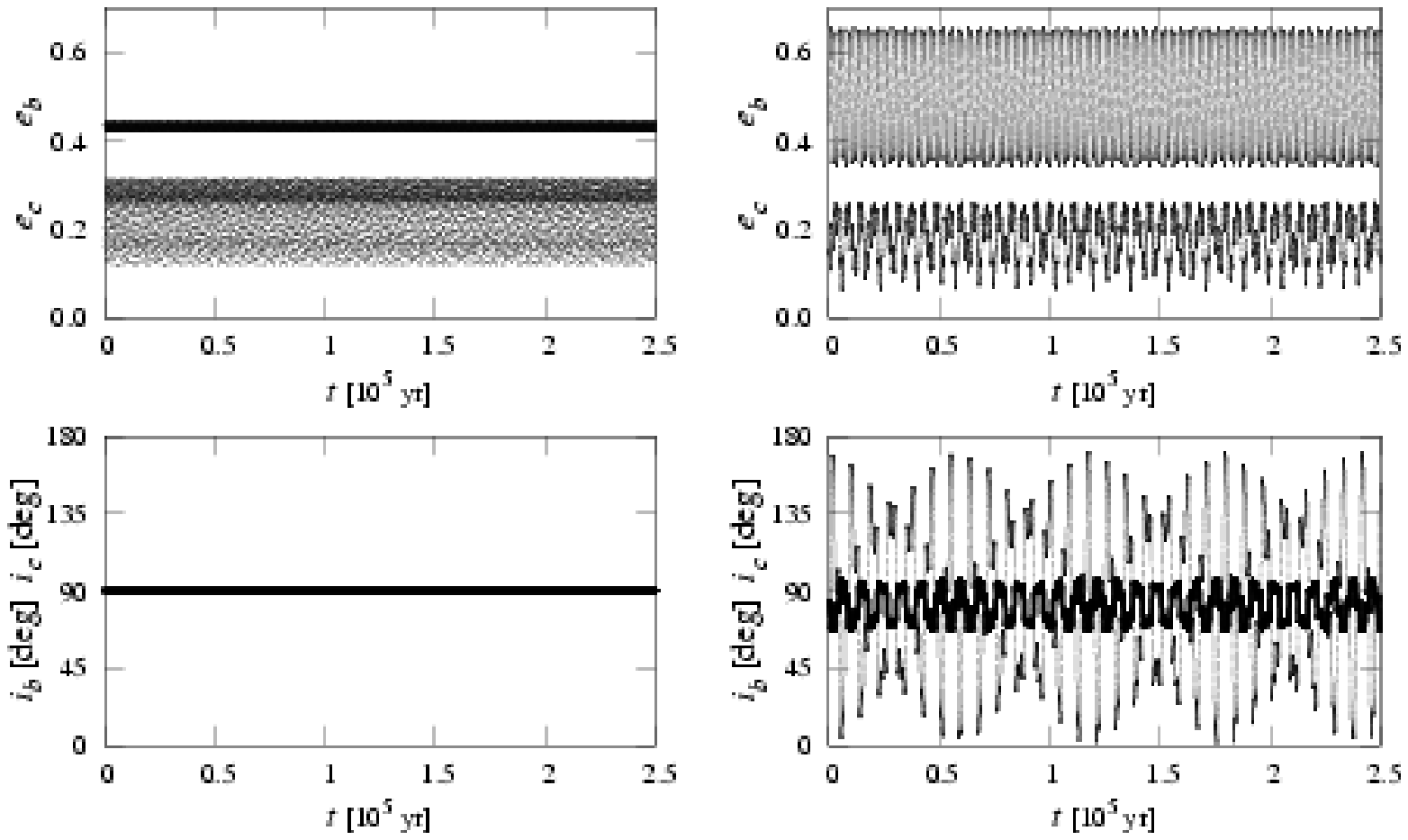}}
   \caption{}
\label{fig:fig5}%
\end{figure*}
%
%
\begin{figure*}
   \centering
   \hbox{\includegraphics[]{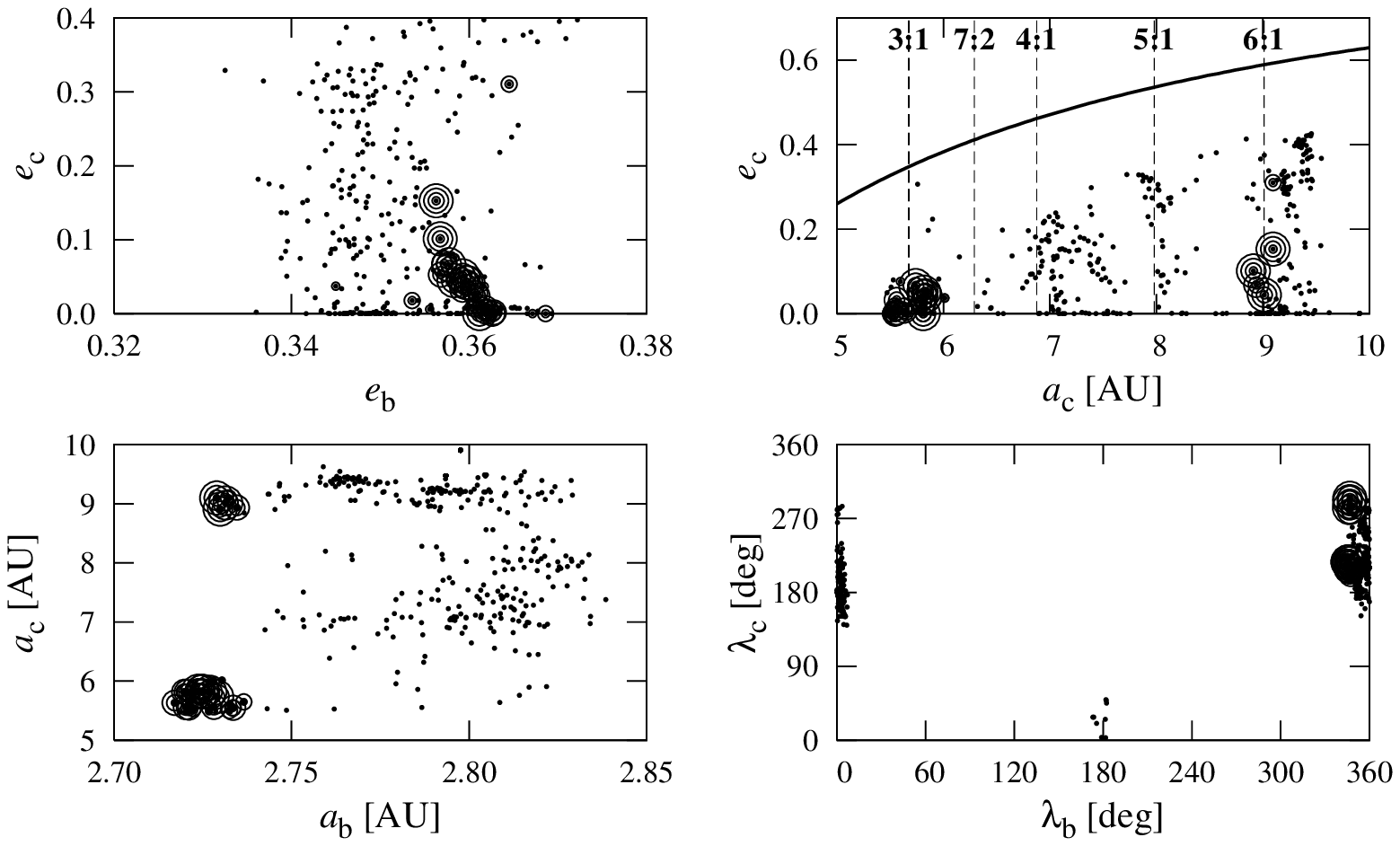}}
   \caption{}
\label{fig:fig6}%
\end{figure*}
%
%
\begin{figure*}
   \centering
   \hbox{\includegraphics[]{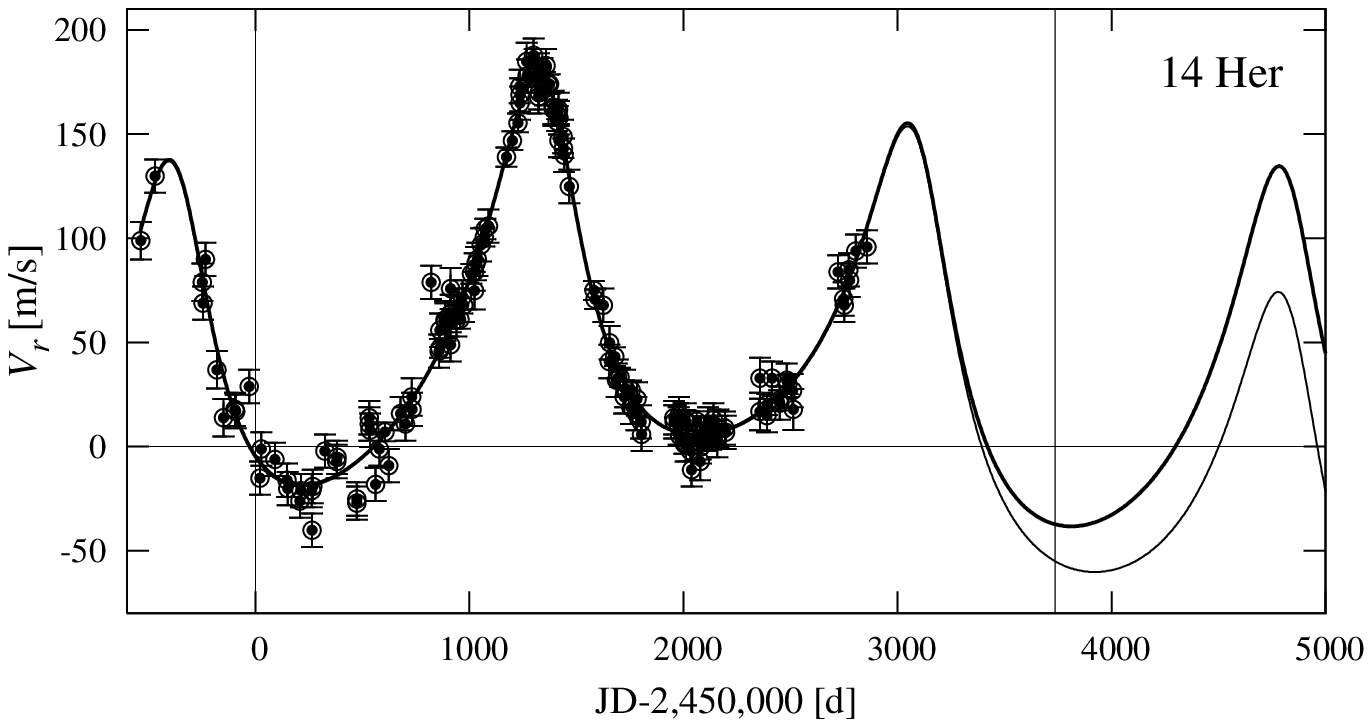}}
   \caption{}
\label{fig:fig7}%
\end{figure*}
%
%
\begin{figure*}
   \centering
   \hbox{\includegraphics[]{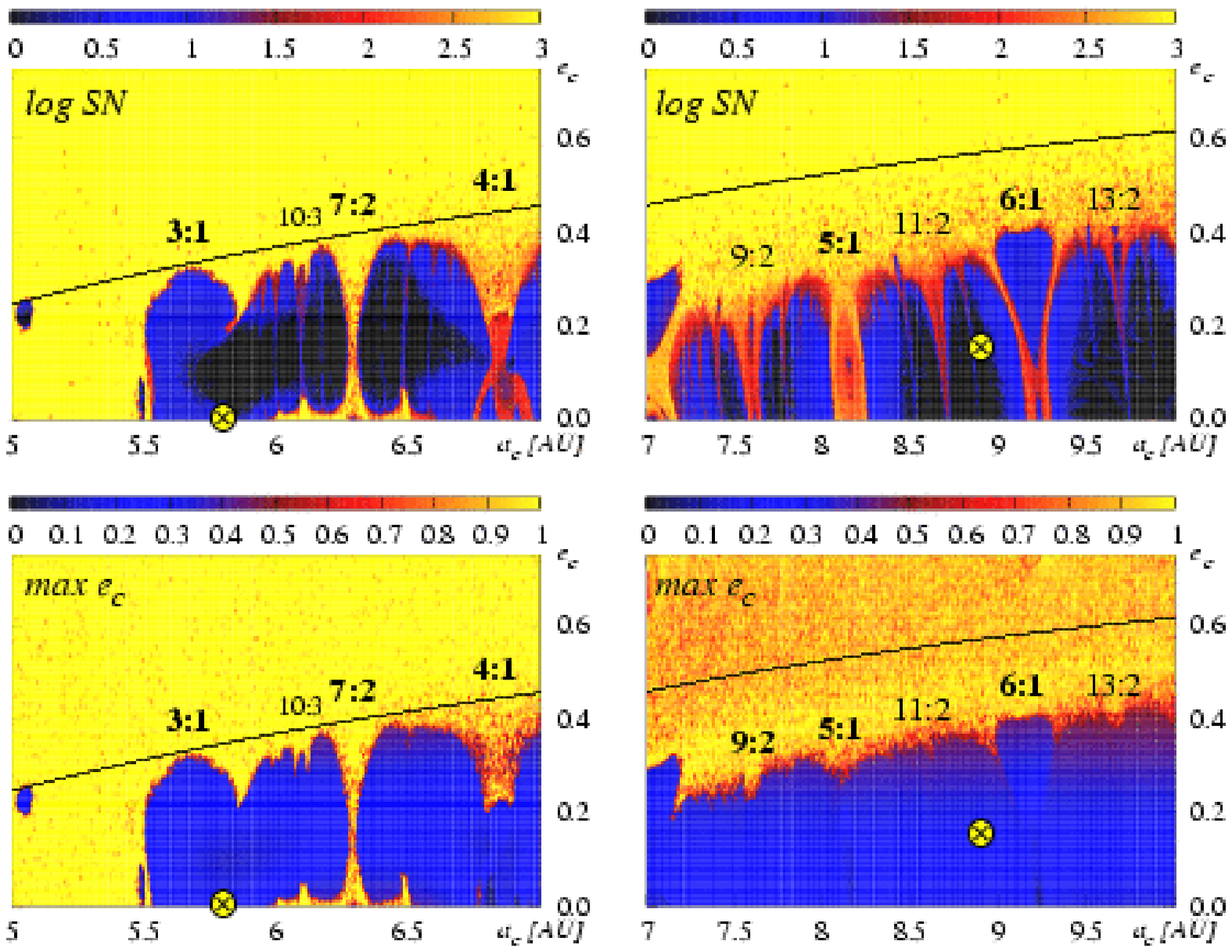}}
   \caption{}
\label{fig:fig8}%
\end{figure*}

%
%

\begin{figure*}
   \centering
   \hbox{\includegraphics[]{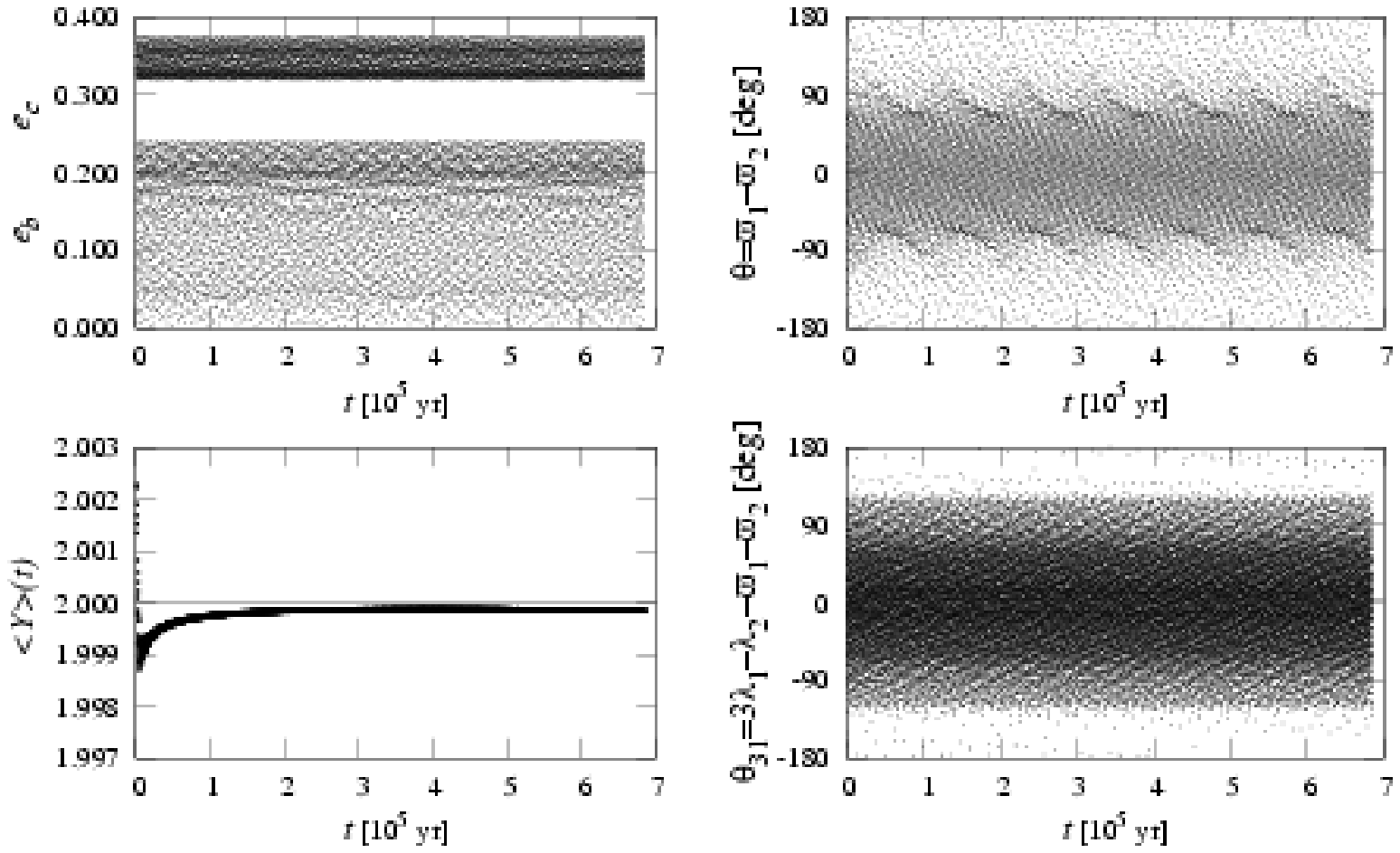}}
   \caption{}
\label{fig:fig9}%
\end{figure*}

%
%
\begin{figure*}[!th]
   \centering
   \hbox{\includegraphics[]{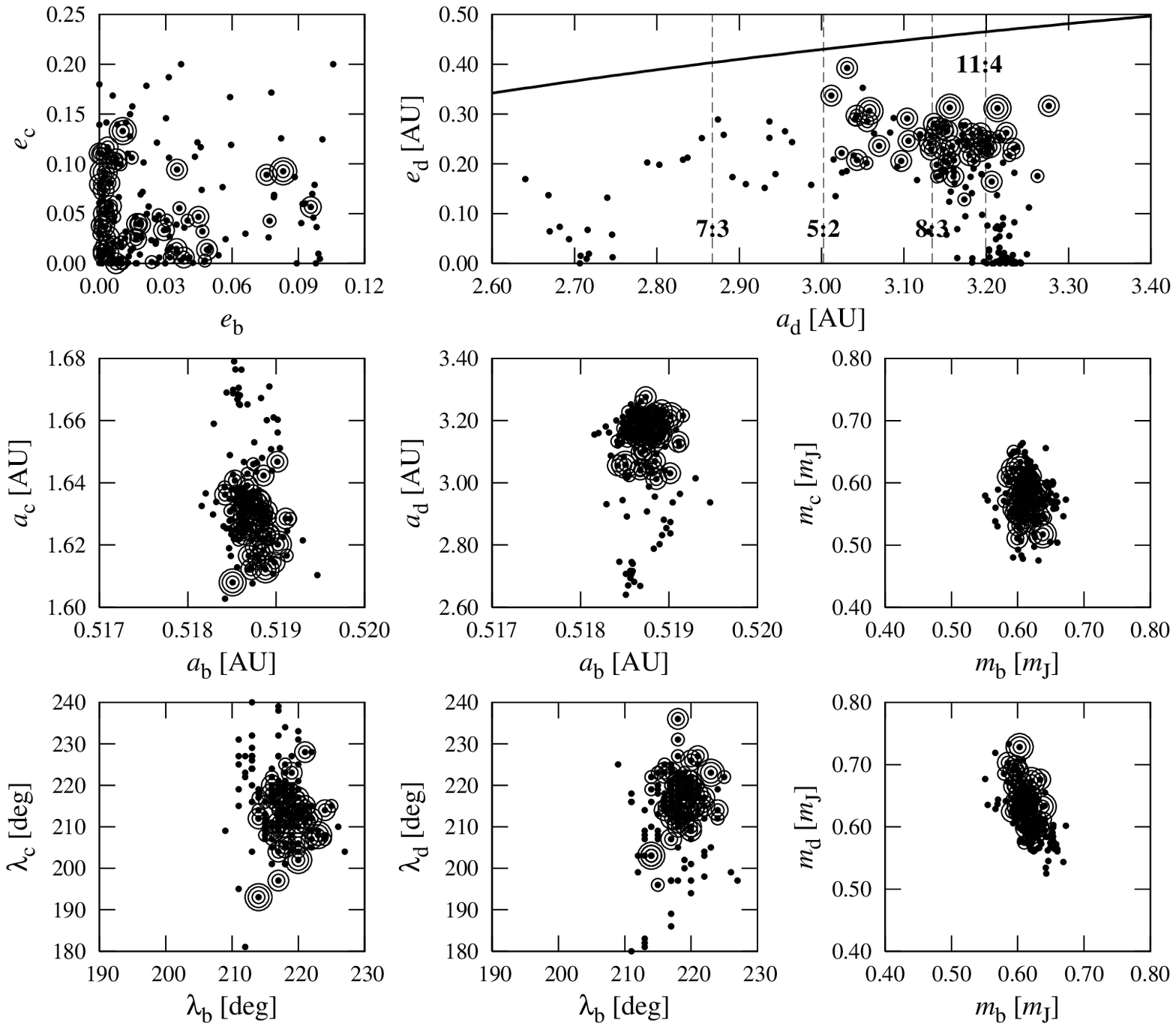}}
\caption{}
\label{fig:fig10}%
\end{figure*}

%
%
\begin{figure*}
   \centering
   \hbox{\includegraphics[]{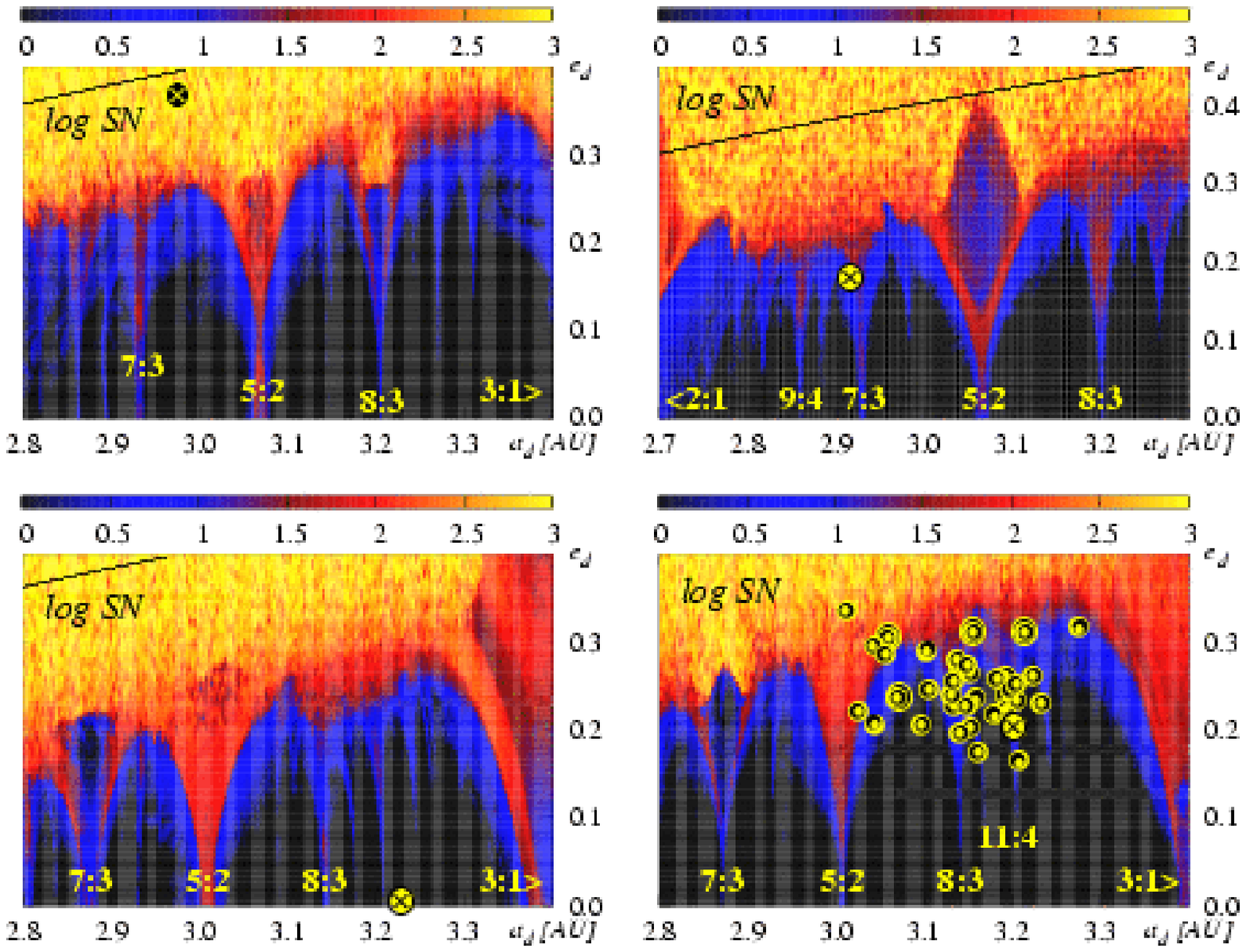}}
   \caption{
   }
\label{fig:fig11}%
\end{figure*}
%
%
\begin{figure*}
   \centering
   \hbox{\includegraphics[]{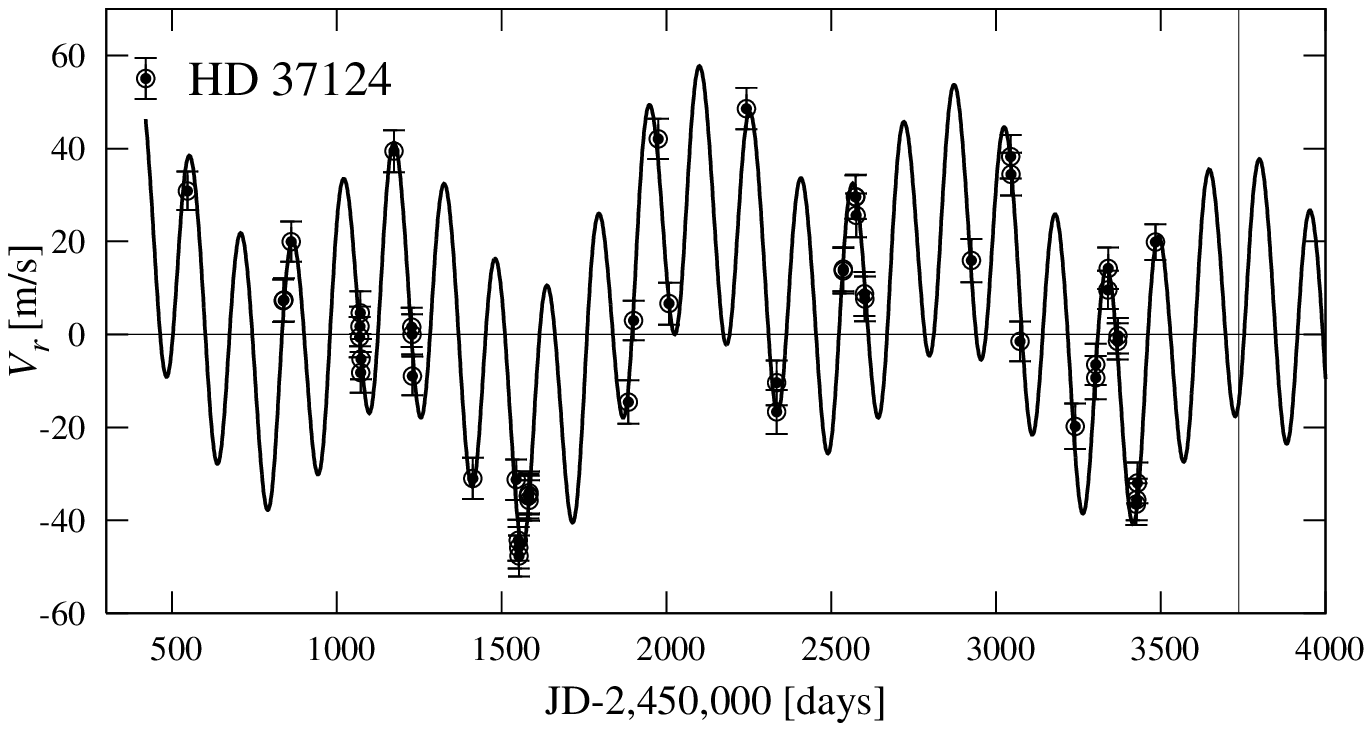}}
\caption{}
\label{fig:fig12}%
\end{figure*}
%
%
%
\begin{figure*}
   \centering
   \hbox{\includegraphics[]{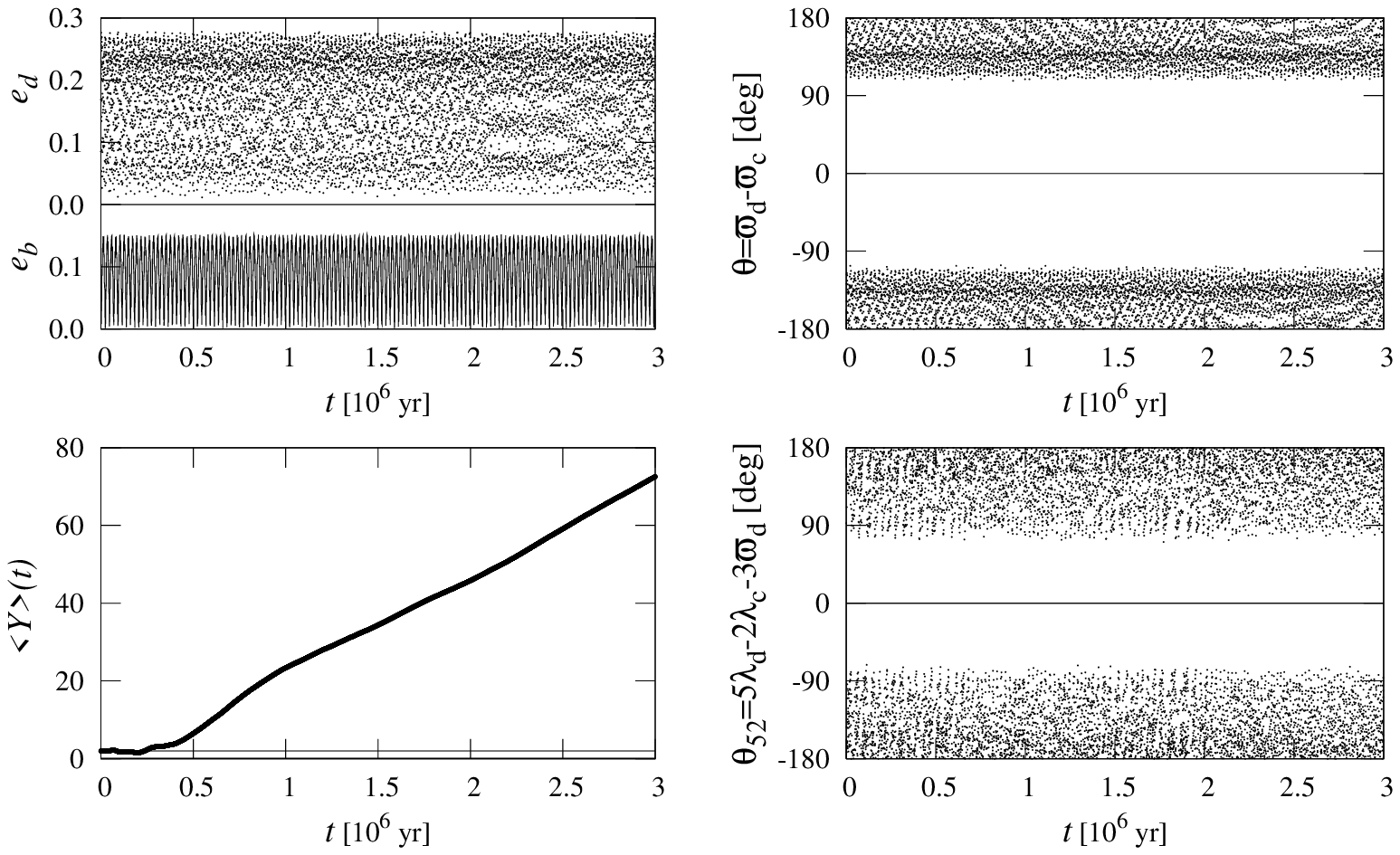}}
\caption{}
\label{fig:fig13}%
\end{figure*}

%
%
\begin{figure*}
   \centering
   \hbox{\includegraphics[]{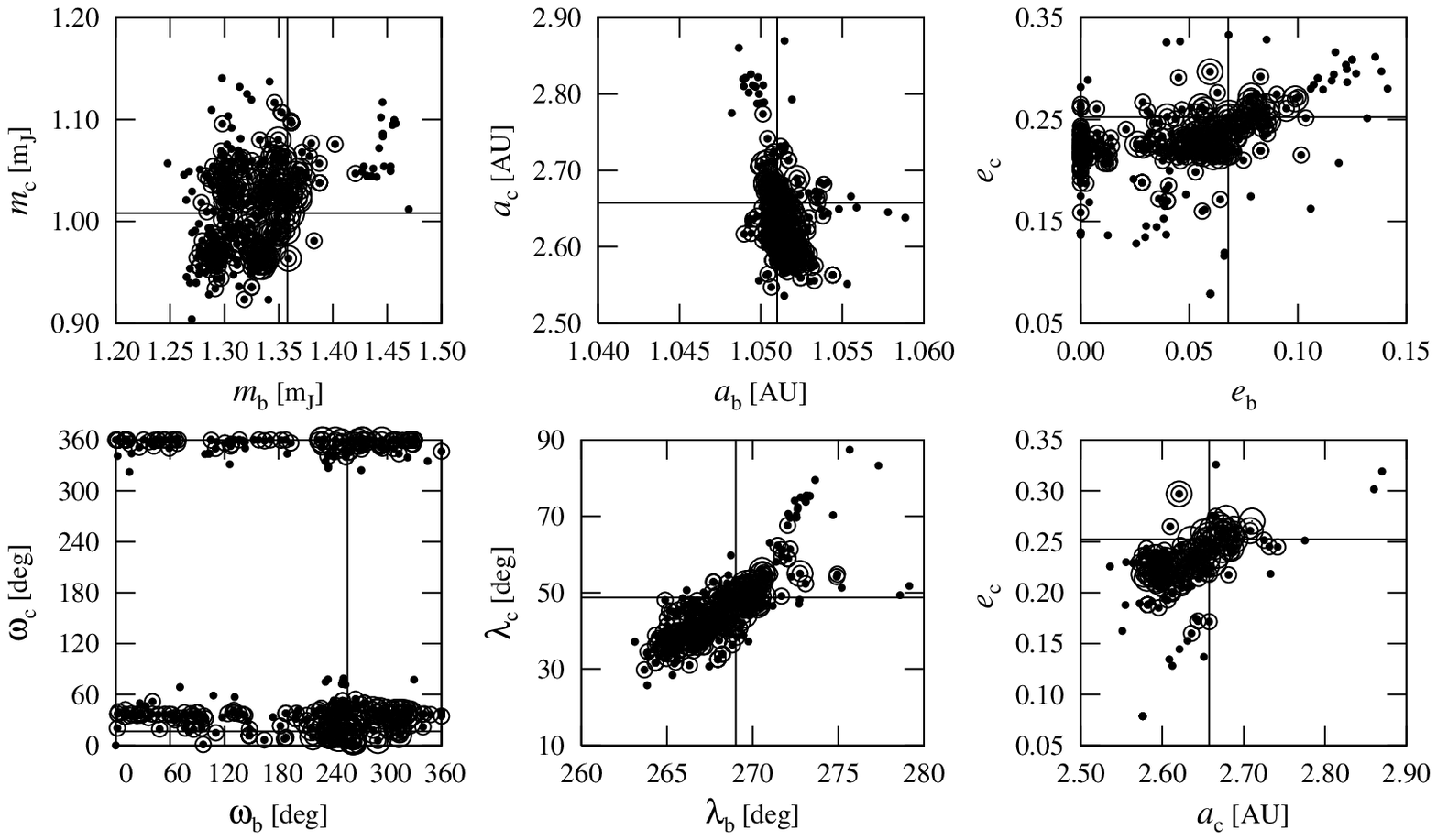}}
   \caption{
   }
\label{fig:fig14}%
\end{figure*}

%
%
\begin{figure*}
   \centering
   \hbox{\includegraphics[]{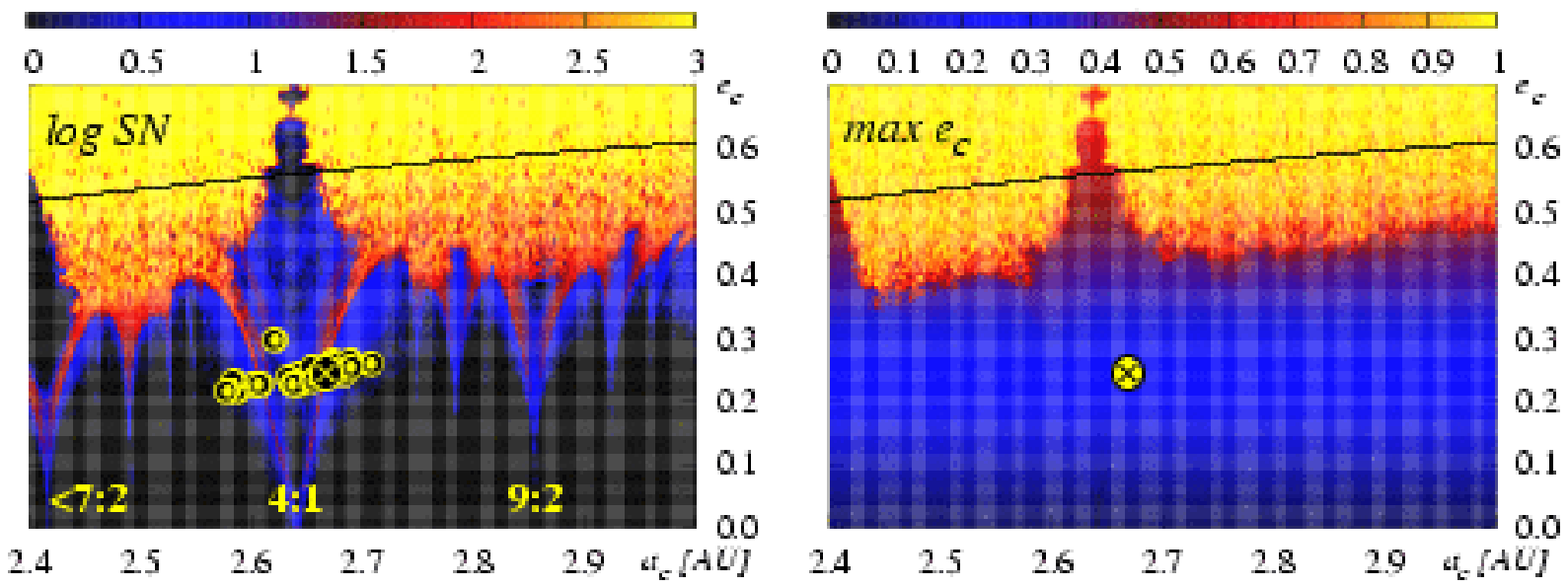}}
   \caption{
   }
\label{fig:fig15}%
\end{figure*}

%
%
\begin{figure*}
   \centering
   \hbox{\includegraphics[]{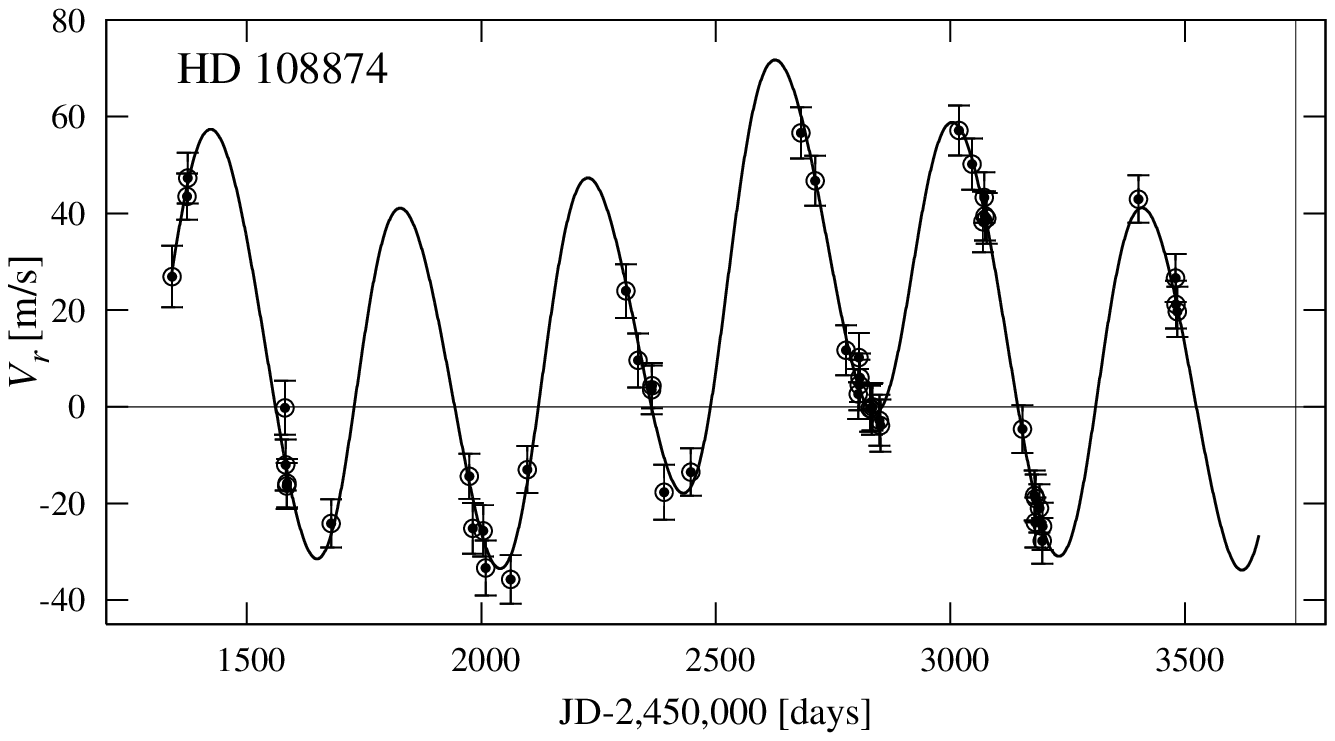}}
   \caption{
   }
\label{fig:fig16}%
\end{figure*}

%
%
\begin{figure*}
   \centering
   \hbox{\includegraphics[]{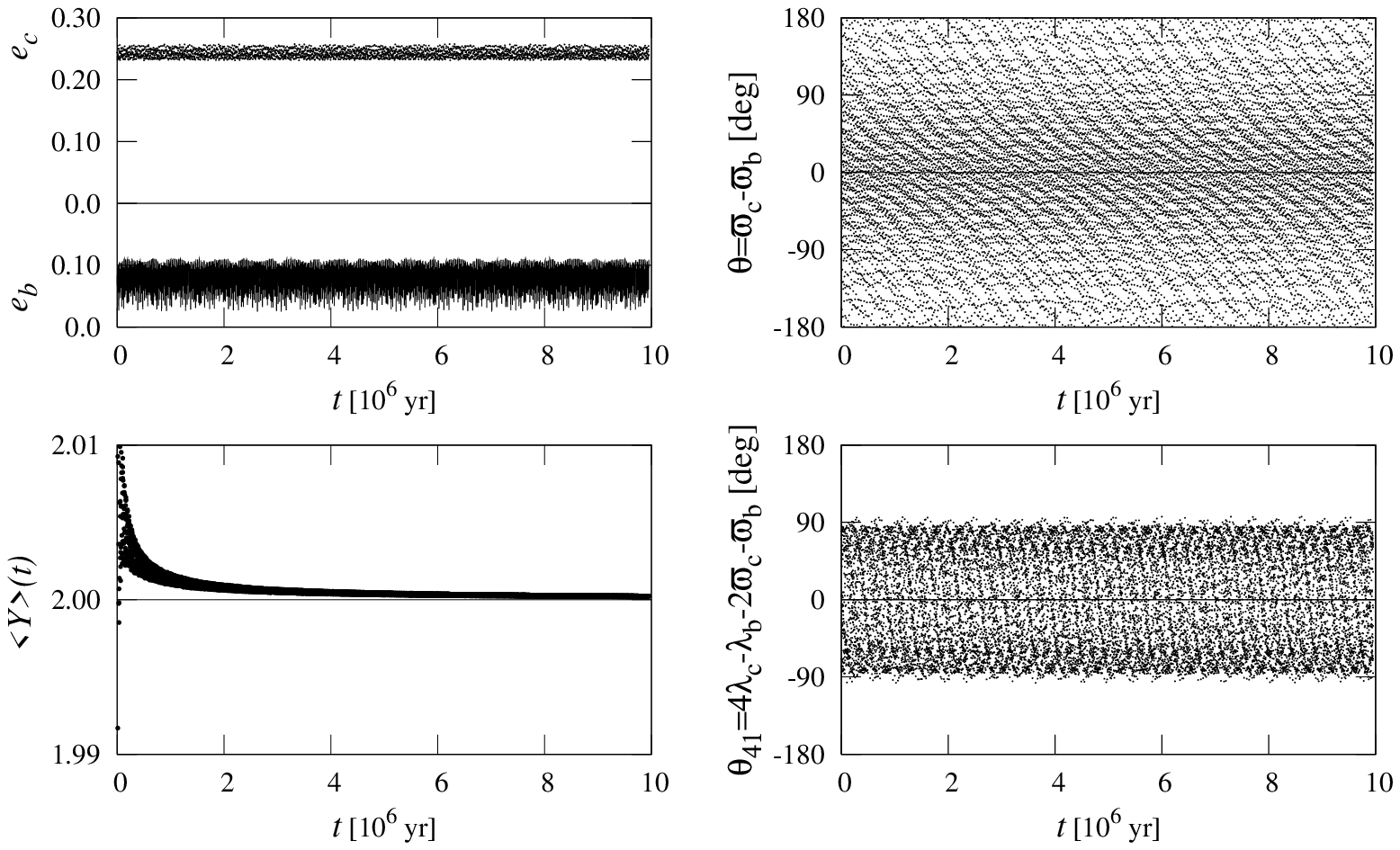}}
   \caption{
   }
\label{fig:fig17}%
\end{figure*}

\end{document}